\documentclass[fleqn,usenatbib]{mnras}
\usepackage{fix-cm}

\usepackage{newtxtext,newtxmath}

\usepackage[T1]{fontenc}

\DeclareRobustCommand{\VAN}[3]{#2}
\let\VANthebibliography\thebibliography
\def\thebibliography{\DeclareRobustCommand{\VAN}[3]{##3}\VANthebibliography}

\usepackage{graphicx}	
\usepackage{amsmath}	
\usepackage{verbatim}
\usepackage{url}

\newcommand{\chg}[1]{#1}

\title[$\gamma$-ray variability of changing-look blazars]{$\gamma$-ray variability of changing-look blazars:
FSRQ-like jet behaviour}

\author[Malik et al.]{
Zahoor Malik,$^{1}$\thanks{E-mail:2030974402@ufs4life.ac.za}
B. van Soelen,$^{1}$\thanks{E-mail:vansoelenb@ufs.ac.za}
Sikandar Akbar$^{2}$\thanks{E-mail:darprince46@gmail.com}
and Zahir Shah$^{3}$\thanks{E-mail:shahzahir4@gmail.com}
\\
$^{1}$Department of Physics, University of the Free State, PO Box 339, Bloemfontein 9300, South Africa\\
$^{2}$Department of Physics, University of Kashmir, Srinagar 190006, India\\
$^{3}$Manipal Centre for Natural Sciences, Centre of Excellence, Manipal Academy of Higher Education, Manipal 576104, India
}

\date{Accepted XXX. Received YYY; in original form ZZZ}

\pubyear{\the\year{}}

\begin{document}
\label{firstpage}
\pagerange{\pageref{firstpage}--\pageref{lastpage}}
\maketitle

\begin{abstract}
Changing-look blazars (CLBs) move between the flat-spectrum radio quasar
(FSRQ) and BL Lacertae (BL Lac) optical classifications. Either the
accretion flow switches regime, or a brightening jet continuum swamps
the broad lines while the system itself is unchanged. The two pictures predict the same optical behaviour but differ in the
jet, so the $\gamma$-ray emission \chg{could potentially separate them.
Catalogue quantities averaged over the mission cannot, because a
permanently intermediate source and one alternating between two states
return the same mean.} We therefore use 17.9\,yr of \textit{Fermi}-LAT
light curves at 3-, 7- and 30-day cadence, comparing CLBs with BL Lac
and FSRQ controls selected and analysed identically. CLBs and FSRQs have
indistinguishable fractional variability, with medians of 0.483 and
0.437 at 30-day cadence ($p = 0.94$), while BL Lacs are far less
variable at 0.262 ($p = 1.4\times10^{-4}$). The result survives
restricting the controls to low-synchrotron-peaked sources and matching
in flux and sampling. Several further diagnostics group CLBs with
FSRQs, as does their Eddington-scaled disc luminosity. Followed through time, CLB
light curves show no excess of bimodal flux distributions, and
the direction of the optical transition does not predict how the
$\gamma$-ray spectrum evolves. CLBs do reach hard, BL Lac-like spectra
more often than FSRQs, but never for longer than chance allows. Only in
this, and in how their spectra respond to brightening, do they fall
between the two classes. The changing-look phenomenon therefore leaves
little imprint on the $\gamma$-ray jet, favouring a variable jet
continuum over a change of accretion regime.
\end{abstract}

\begin{keywords}
galaxies: active -- BL Lacertae objects: general -- quasars: general --
gamma-rays: galaxies -- methods: statistical
\end{keywords}


\section{Introduction}
\label{sec:intro}

Blazars are radio-loud active galactic nuclei whose relativistic jets are
aligned close to the line of sight \citep{BlandfordRees1978,
UrryPadovani1995}. The alignment boosts the jet emission, which then
dominates the observed spectral energy distribution from radio to
$\gamma$-rays \citep{Blandford2019, MadejskiSikora2016}. Blazars are
strongly variable at every wavelength, and this variability is one of the
principal tools available for probing physical conditions in the jet
\citep{Ulrich1997, Abdo2010}.

Blazars are traditionally separated into two classes on the basis of
their optical spectra. FSRQs show broad emission lines with \chg{an
absolute} rest-frame equivalent width \chg{$|EW| > 5$\,\AA}, while BL Lac objects show weak lines or
none at all \citep{Stickel1991, Stocke1991, UrryPadovani1995}. The
5\,\AA\ threshold is a historical convention rather than a physical
boundary, but the division has proved durable because it correlates with
other properties. FSRQs are more luminous, have softer $\gamma$-ray
spectra and show stronger long-term variability, while BL Lacs are
fainter, spectrally harder and often peak at higher synchrotron
frequencies \citep{Fossati1998, Ghisellini1998, Ackermann2015,
Ajello2020}.

A physical interpretation of the division was given by
\citet{Ghisellini2011}, who argued that it reflects the accretion
regime. FSRQs accrete efficiently, producing a luminous accretion disc
that photoionises the broad-line region and supplies external photons
for inverse-Compton scattering. BL Lacs host radiatively inefficient
flows in which the broad-line region is weakly illuminated or absent
\citep{Sbarrato2012, Ghisellini2014}. The transition between the two
regimes is expected near $L_{\rm disk}/L_{\rm Edd} \sim 10^{-2}$, and
this framework accounts naturally for the observed correlation between
line strength and Compton dominance \citep{Paliya2021}.

Against this background, a growing number of blazars have been found to
cross the classification boundary. Early individual cases
\citep{Vermeulen1995, Bianchin2009, Ghisellini2011} have been joined by
systematic searches in archival and survey spectroscopy
\citep{Shaw2012, Ruan2014, AlvarezCrespo2016, PenaHerazo2021,
Foschini2021, Foschini2022, Xiao2022}, and by detailed monitoring of
individual objects \citep{Mishra2021, Ren2024a, Ren2024b}. These
changing-look, or transitional, blazars are usually interpreted as
systems close to the boundary between the two accretion regimes
\citep{Kang2024}, in the same spirit as the changing-look phenomenon in
the wider AGN population \citep{LaMassa2015, Ruan2016,
RicciTrakhtenbrot2023}.

An alternative explanation has been available since the class was first
recognised. \citet{BlandfordRees1978} noted that a Doppler-boosted
continuum can swamp emission lines and make a line-emitting source
appear featureless. \chg{In this picture,} the accretion flow does not change
at all. The jet brightens, the lines are buried, and the optical
classification changes while the underlying system does not.
\chg{The two pictures are not mutually exclusive in every detail. If the
jet is coupled to the accretion flow, an increase in the accretion rate
would brighten both the broad lines and the jet, and the lines would still
be swamped provided the jet continuum rises by a larger factor than the
line flux. The converse does not hold: the jet can brighten with no change
in accretion at all, because its observed output is governed by processes
within the jet itself, such as the passage of shocks \citep{MarscherGear1985},
magnetic reconnection \citep{Sironi2015} or a change in the Doppler factor
of the emitting region \citep{Raiteri2017}, which operate on timescales of
days to months, shorter than the timescales, typically years, over which
changes of accretion state are observed in changing-look AGN
\citep{RicciTrakhtenbrot2023}.}
\citet{Ruan2014} favoured this interpretation for the transitional
blazars they identified, and related arguments have been made for the
so-called masquerading BL Lacs \citep{Giommi2012, Giommi2013,
Padovani2019}, whose optical spectra are diluted by jet emission. The
anti-correlation between equivalent width and continuum luminosity seen
across the blazar population \citep{Xiao2022} follows naturally from the
same effect.

Separating the two pictures is not straightforward, because they predict
the same optical behaviour. They differ, however, in what should happen
in the jet. If the accretion regime changes, the jet is expected to
change with it, since the external photon field that dominates the
high-energy emission of FSRQs is supplied by the accretion structures
\citep{Sikora1994, DermerSchlickeiser1993}. If the lines are instead
swamped, the jet need not change at all. The $\gamma$-ray emission,
which is produced in the jet, therefore offers a way to distinguish
them.

Population studies of CLBs have so far used catalogue quantities.
\citet{Kang2024} applied model-based clustering to
$\gamma$-ray and accretion-related properties of 105 CLBs and found
them to lie between the two classes, with an Eddington ratio close to
the value expected for a change of accretion mode. \citet{Zhao2026} reached a similar conclusion from the
$\gamma$-ray photon spectral index, applying a Gaussian mixture model
to 3934 jet-dominated AGN from the fourth LAT AGN catalogue and identifying
a population of transition sources at intermediate $\Gamma_\gamma$.
In our own earlier work \citep{Malik2026} we compared 109 CLBs with
confirmed BL Lacs and FSRQs using 4FGL-DR4 spectral and variability
parameters \chg{\citep{Abdollahi2020, Ballet2023}} together with X-ray properties from the Living Swift-XRT
Point Source catalogue \chg{\citep[LSXPS;][]{Evans2023}}. Principal component analysis, uniform manifold
approximation and projection, and a random-forest classifier all placed
the CLB population in the region between the two classes, with its
centroid closer to the FSRQs, and the X-ray properties showed the same
behaviour. That study, like those above, used quantities averaged over
the mission.

These analyses share a limitation that comes from the data they use.
Catalogue parameters are averages over the whole mission, and an average
does not readily distinguish a source that is permanently intermediate
from one that spends part of its life in each of two states. Both return the
same value. To tell them apart, the time axis that the averaging removes
has to be restored.

That is what we do here, using the light curves themselves. The
\textit{Fermi}-LAT Light Curve Repository \citep[LCR;][]{Abdollahi2023}
provides uniformly analysed light curves at three cadences for a large
sample of variable $\gamma$-ray sources, with a photon index measured in
every time bin. This allows the variability properties of CLBs to be
characterised directly and compared with control populations selected
and analysed in the same way. Section~\ref{sec:data} describes the
samples and the data. Sections~\ref{sec:fvar}--\ref{sec:time} present
the flux variability, the spectral variability, the flux and index
distributions and power spectra, the photon-index/flux relation, and the
time-resolved behaviour. Section~\ref{sec:engine} compares the
central-engine properties and Section~\ref{sec:discussion} discusses
what the results imply. We adopt a flat cosmology with
$H_{0} = 70$\,km\,s$^{-1}$\,Mpc$^{-1}$, $\Omega_{\rm m} = 0.3$ and
$\Omega_{\Lambda} = 0.7$.

\section{Sample and data}
\label{sec:data}

\subsection{The changing-look sample}
\label{sec:clbsample}

We take our CLB sample from the changing-look blazar catalogue compiled
by \citet{Kang2023}, which gathers reports from the literature into a
single list. The catalogue is deliberately inclusive. Its 508 entries,
covering 316 distinct sources, range from objects with
spectroscopically observed transitions to candidates identified by
classifiers that have not yet been observed to change. Because the
strength of the evidence differs between these categories, we assign
each source to one of three tiers according to how it was identified.

Tier~A contains 52 sources for which a transition has been observed and
its direction recorded. These come from \citet{Xiao2022}, who compared
the broad-line equivalent widths compiled by \citet{Paliya2021} with
the optical classification listed in 4LAC-DR2 \citep{Ajello2020}. Of these, 42 sources
move from a BL Lac-like to an FSRQ-like spectrum ($B\!\rightarrow\!F$)
and 10 in the opposite sense ($F\!\rightarrow\!B$). Tier~B contains 59
sources that earlier studies report as changing-look or transitional on
the basis of repeat spectroscopy, but for which the direction of the
change is not recorded in the compilation \citep{Vermeulen1995,
Ghisellini2011, Shaw2012, Ruan2014, Cutini2014, AlvarezCrespo2016,
PashamWevers2019, Padovani2019, PenaHerazo2021, Foschini2021,
Foschini2022, Mishra2021}. Tier~C contains 205 sources identified by
classifier output or by indirect diagnostics alone. We use tiers~A
and~B, 111 sources in total, as the primary sample, and set tier~C
aside. Results for tier~A alone are quoted where the smaller sample
permits.

The sample was cross-matched with 4FGL-DR4 \citep{Abdollahi2020,
Ballet2023} first by 4FGL designation and, where that failed, by the
name of the low-frequency counterpart. The second step is needed
because 4FGL designations encode source position, so a source whose
position is revised between catalogue releases acquires a new name.
OQ~334, the best-studied CLB in the literature and the source for which
\citet{Ren2024b} proposed their $\gamma$-ray state criterion,
moved by 183~arcsec between the release used by \citet{Kang2023} and
DR4, and is recovered only by counterpart matching.

\subsection{Control samples}
\label{sec:controls}

Control samples were drawn from 4FGL-DR4 using the class1 field,
giving 1427 BL Lacs and 775 FSRQs once tier A and B CLBs are
removed. Tier C candidates are retained in the controls and flagged;
200 of them fall in the BL Lac control and four in the FSRQ control,
while the remaining source is classified as a blazar candidate of
uncertain type and so enters neither. Removing them would
preferentially remove the most FSRQ-like members of the BL Lac sample,
and would tend to enlarge any difference we subsequently report, so
retaining them is the more conservative choice.

\subsection{Light curves}
\label{sec:lightcurves}

Light curves were obtained from the LCR \citep{Abdollahi2023} at 3-, 7-
and 30-day cadence, using photon fluxes with the photon index left free
and a detection threshold of $TS \ge 4$. \chg{The retrieved data span MJD
54698--61238, that is 17.9\,yr.} The LCR performs the likelihood
analysis uniformly for all sources, which removes a possible systematic
difference between samples that would otherwise be difficult to control.
The two control samples were queried together with all 316 sources of the changing-look catalogue, not only the 111 of tiers A and B, giving 2314 distinct sources. Ten of these were renamed between the DR2 and DR4 releases, and the LCR files them under their earlier designations, so they were retrieved under those names. Of the 2314 sources, 1069 returned data.
The remainder are absent from the repository, for two reasons. \chg{First,
the LCR} was built from the
4FGL-DR2 source list \citep{Ballet2020}, so sources first catalogued in
DR3 or DR4 are not included at all. \chg{Second, the LCR} restricts itself to
sources whose DR2 variability index exceeds 21.67, that is, those with
less than a 1 per cent probability of being steady. Our samples are
selected from DR4, so this coverage difference is a property of the
repository rather than of our selection.

The coverage criterion is worth noting explicitly. Because the LCR
contains only sources already known to be variable, the BL Lac control
is biased towards its more variable members. Any difference we report
between CLBs and BL Lacs is therefore conservative in this respect.

We retain time bins with $TS \ge 4$, $F/\sigma_{F} > 2$,
$\Gamma/\sigma_{\Gamma} > 2$, $\Gamma < 6$ and a converged fit. These
follow \citet{Shah2025} so that our measurements can be compared
directly with published values for the canonical classes. Two further
cuts are required by the data format. The LCR stores flux and photon
index uncertainties as interval bounds rather than as standard
deviations, and the two arrays are ordered differently, so half the
absolute interval width is used in both cases. In addition, a small
number of bins report a vanishing uncertainty together with an
unphysically large flux; these survive a signal-to-noise cut precisely
because their nominal uncertainty is zero. We therefore reject bins with
non-positive uncertainty, with flux above
$10^{-3}$\,ph\,cm$^{-2}$\,s$^{-1}$, or with flux exceeding 20 times the
source median while remaining below $TS = 100$.

\chg{Because the LCR supplies the $TS$ of every bin, we also repeated the
main comparisons with stricter thresholds of $TS \ge 9$ and $TS \ge 25$
imposed on top of these cuts; the outcome is noted in the relevant
sections below.}

After all cuts, the 30-day sample contains 64 CLBs, 204 BL Lacs and 389
FSRQs with at least 20 usable bins; 32 of the CLBs belong to tier~A. The
corresponding numbers are 65, 169 and 429 at 7-day cadence and 68, 169
and 462 at 3-day cadence.

\subsection{Choice of cadence}
\label{sec:cadence}

The three cadences are not equally useful for every measurement, and it
is worth setting out at the start which is used where and why.

The 30-day light curves are the primary data set. Each bin accumulates
the most exposure, so the largest number of bins pass the significance
cuts and the per-bin photon index is best determined. At 30-day cadence
the median \chg{per-bin uncertainty on the photon index $\Gamma$ (taking the
median over bins for each source and then over sources)} is \chg{0.25} for CLBs, 0.26 for BL Lacs and
\chg{0.29} for FSRQs; at 3-day it rises to 0.38, 0.42 and 0.43. The control samples are also
largest at 30-day cadence, since faint BL Lacs are frequently undetected
in short bins.

The 7- and 3-day light curves are used where the question concerns
behaviour on shorter timescales, and as consistency checks on the
30-day results. We report the fractional variability, the spectral
variability, the photon-index/flux correlation and the
distribution-model comparison at all three cadences. \chg{The
distribution models, however, are best determined at 30-day cadence: the
two-component model has three more free parameters than the single
lognormal, and each source contributes fewer usable bins at shorter
cadence (Section~\ref{sec:models}).} The power spectra are reported at
30-day cadence only, \chg{because too few CLBs have light curves long
enough for a segmented periodogram at 7- and 3-day cadence
(Section~\ref{sec:psd}).}

\subsection{Consistency checks}
\label{sec:validation}

Two checks were made before the analysis, both of which test the
reliability of the quantities the rest of the paper depends on.

The first concerns the per-bin photon index uncertainties. These enter
every noise-debiased quantity we compute, so an error in reading them
would propagate throughout. For a likelihood fit the uncertainty should
scale as $TS^{-1/2}$, and we verified that it does, using every
bin with a measured index from all sources that returned data\chg{,
56\,735 bins at 30-day cadence. Binning these in $TS$, the median
$\sigma_{\Gamma}$ falls from 0.55 for $4 \le TS < 10$ to 0.02 above
$TS = 5000$, and a power law fitted to the binned medians across the full
range has a slope of $-0.46$, close to the expected $-1/2$; the 7- and
3-day light curves give $-0.47$ and $-0.46$.} This confirms both that the uncertainties are read correctly from
the repository and that they behave as counting statistics require.
\chg{They can therefore be subtracted as the noise term in the
fractional variability $F_{\rm var}$ and the intrinsic spectral
variability $\sigma_{\Gamma,\rm int}$ (equations~\ref{eq:fvar}
and~\ref{eq:sigint}).}

The second concerns the absolute scale of the variability measurements.
Since our results rest on comparing samples, a systematic error common
to all of them would not be visible internally. We therefore recomputed
the fractional variability for six bright blazars for which published
values exist \citep{Shah2025}. Our values reproduce the published ones,
which indicates that our cuts and our treatment of the
uncertainties agree with an independent implementation.

All probabilities quoted in this paper come from two-sample
Kolmogorov--Smirnov (KS) tests unless stated otherwise.

\section{Flux variability}
\label{sec:fvar}

The fractional variability amplitude \citep{Vaughan2003},
\begin{equation}
F_{\rm var} =
  \frac{\sqrt{S^{2} - \overline{\sigma_{\rm err}^{2}}}}
       {\overline{F}},
\label{eq:fvar}
\end{equation}
measures the intrinsic variability of a light curve once the
contribution of measurement noise has been removed. Here $S^{2}$ is the
sample variance, $\overline{\sigma_{\rm err}^{2}}$ the mean square
uncertainty and $\overline{F}$ the mean flux. The uncertainty on
$F_{\rm var}$ is determined using the equation
\begin{equation}
\sigma_{F_{\rm var}} =
  \sqrt{\left(\sqrt{\frac{1}{2N}}\,
        \frac{\overline{\sigma_{\rm err}^{2}}}
             {F_{\rm var}\overline{F}^{2}}\right)^{2}
      + \left(\sqrt{\frac{\overline{\sigma_{\rm err}^{2}}}{N}}\,
        \frac{1}{\overline{F}}\right)^{2}},
\label{eq:fvarerr}
\end{equation}
where $N$ is the number of time bins in the light curve
\citep{Vaughan2003}. Because the noise term is
subtracted explicitly, $F_{\rm var}$ can in principle be compared
between sources of differing brightness. In practice a residual
dependence on brightness can remain when few bins survive the quality
cuts, and we test for this in Section~\ref{sec:matched}.

\begin{figure*}
\centering
\includegraphics[width=\textwidth]{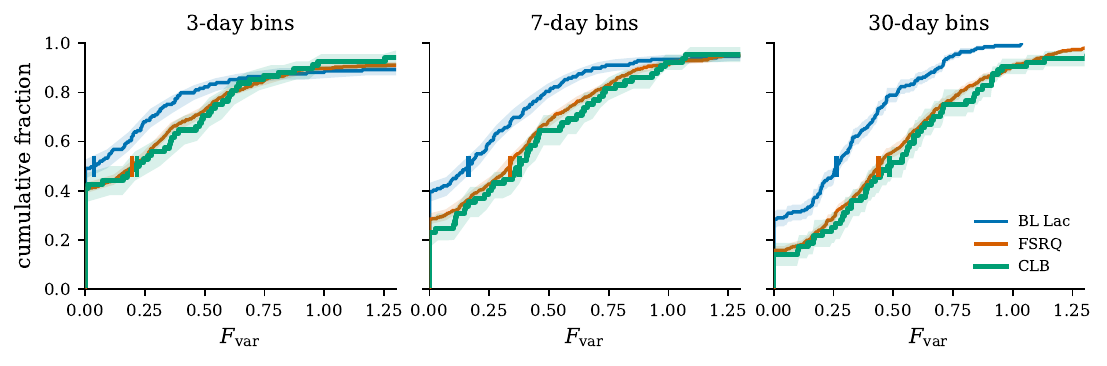}
\caption{Cumulative distributions of the fractional variability
amplitude $F_{\rm var}$ for the three samples, at 3-day (left), 7-day
(middle) and 30-day (right) cadence. Blue denotes BL Lacs, orange FSRQs
and green CLBs. The number of sources contributing to each curve is
given in Section~\ref{sec:lightcurves}. The shaded band around each curve is the 68 per
cent confidence region obtained by bootstrap resampling of the sources,
and indicates how much the curve would move under a different draw from
the same population. The short vertical tick on each curve marks the
median. \chg{The KS probabilities for the CLB--FSRQ and CLB--BL Lac
comparisons at each cadence are listed in Table~\ref{tab:fvar}.} At every cadence
the CLB and FSRQ curves lie close together, while the BL Lac curve is
displaced towards lower variability, that is, it rises earlier. Note
that \chg{both axes are the same in all three panels; in particular the
horizontal ($F_{\rm var}$) axis spans 0--1.3 throughout,} so the
leftward shift of all three curves at shorter cadence reflects the
smaller variability amplitude measured on shorter timescales.}
\label{fig:fvar_ecdf}
\end{figure*}

Fig.~\ref{fig:fvar_ecdf} shows the cumulative distributions at all three
cadences and Table~\ref{tab:fvar} lists the medians. At 30-day cadence
the CLB and FSRQ distributions are statistically indistinguishable
($p = 0.94$), while CLBs and BL Lacs differ at
$p = 1.4 \times 10^{-4}$. The median fractional variability of CLBs,
0.483, is close to that of FSRQs, 0.437, and well above that of
BL Lacs, 0.262. \chg{The result does not depend on the detection
threshold: with $TS \ge 9$ or $TS \ge 25$ imposed on every bin, CLBs and
FSRQs remain indistinguishable ($p = 0.97$ and $0.87$) while CLBs and
BL Lacs differ at $p = 4.2\times10^{-6}$ and $1.1\times10^{-5}$.}

\begin{table*}
\centering
\caption{Median fractional variability by sample and cadence, the number
of CLBs contributing, and the KS \chg{probabilities that the CLB
distribution shares a parent population with the FSRQ
($p_{\rm CLB-FSRQ}$) and BL Lac ($p_{\rm CLB-BLL}$) distributions.} The values in brackets are the median per-source
uncertainty on $F_{\rm var}$ for that sample and cadence, obtained from
equation~(\ref{eq:fvarerr}).}
\label{tab:fvar}
\begin{tabular}{lcccccc}
\hline
Cadence & CLB & FSRQ & BL Lac & $N_{\rm CLB}$ & \chg{$p_{\rm CLB-FSRQ}$} & $p_{\rm CLB-BLL}$\\
\hline
30\,d & 0.483 (0.027) & 0.437 (0.036) & 0.262 (0.056) & 64 & \chg{0.94} & $1.4\times10^{-4}$\\
7\,d  & 0.375 (0.029) & 0.336 (0.040) & 0.164 (0.073) & 65 & \chg{0.80} & $5.5\times10^{-3}$\\
3\,d  & 0.216 (0.029) & 0.196 (0.045) & 0.036 (0.081) & 68 & \chg{0.88} & $5.8\times10^{-2}$\\
\hline
\end{tabular}
\end{table*}

\begin{figure}
\centering
\includegraphics[width=\columnwidth]{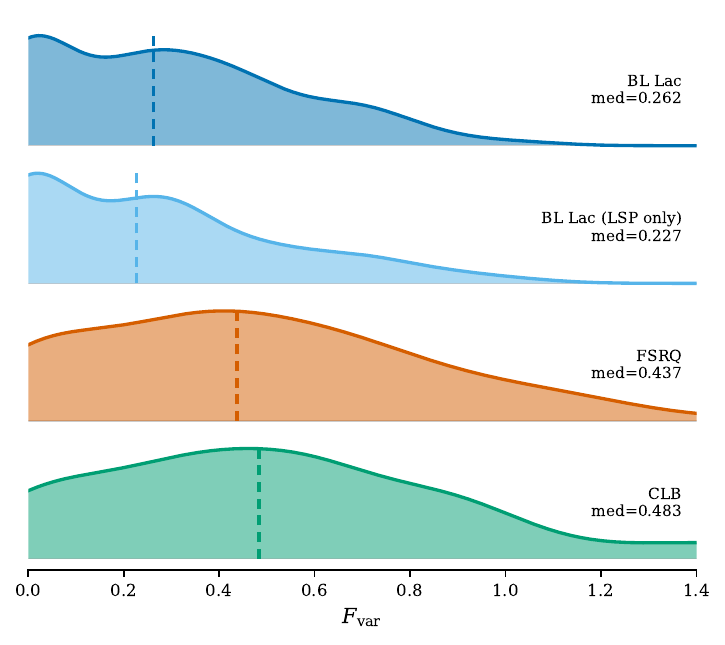}
\caption{Distribution of $F_{\rm var}$ at 30-day cadence, shown as
kernel density estimates offset vertically for clarity. Each curve is
normalised to the same peak height, so the vertical extent carries no
information and only the horizontal position and width should be
compared. From top to
bottom: the full BL Lac control, the BL Lac control restricted to
low-synchrotron-peaked sources, the FSRQ control and the CLB sample. The
dashed vertical line within each curve marks the median. Restricting the
BL Lac control to LSP sources shifts its median slightly lower rather
than towards the CLB value, which shows that the CLB--BL Lac difference
is not produced by the different synchrotron peak distributions of the
two samples.}
\label{fig:ridgeline}
\end{figure}

Fig.~\ref{fig:ridgeline} shows the same 30-day distributions as
densities rather than cumulatively, which makes the relative positions
of the four curves easier to compare directly; the second row is
discussed in Section~\ref{sec:matched}.

The same ordering holds at 7- and 3-day cadence. The CLB--BL Lac
difference remains significant at 7-day cadence
($p = 5.5 \times 10^{-3}$) but not at 3-day
($p = 5.8 \times 10^{-2}$). This appears to be a question of
sample size rather than \chg{to have a physical reason: faint BL Lacs are}
frequently undetected in short time bins, so the usable control shrinks. At 3-day cadence it falls to 169 sources, and its median $F_{\rm var}$
of 0.036 lies well below the median per-source uncertainty of 0.081
quoted in Table~\ref{tab:fvar}, so most of these light curves are close
to the noise floor once the debiasing has been applied. We
therefore treat the 30-day results as primary and the shorter cadences
as consistency checks, and note that the ordering of the three samples
is the same in all cases.

\begin{figure}
\centering
\includegraphics[width=\columnwidth]{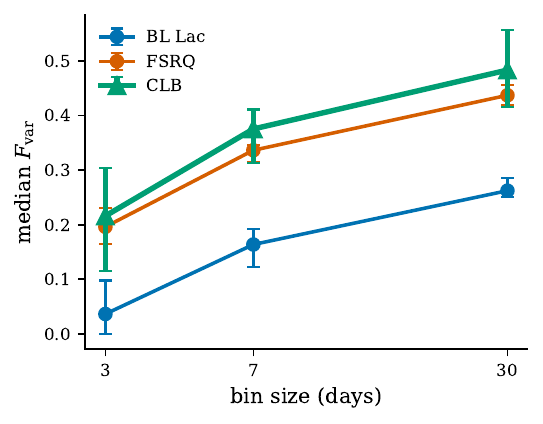}
\caption{Median fractional variability as a function of bin size, on a
logarithmic horizontal scale. Error bars give the 68 per cent confidence
interval on the median obtained by bootstrap resampling. Triangles denote CLBs, circles the two
control samples. All three samples show larger amplitudes at longer
binning, as expected when the light curves are dominated by red noise,
and the CLB and FSRQ points track one another across the full range
while the BL Lac points remain systematically lower.}
\label{fig:timescale}
\end{figure}

Fig.~\ref{fig:timescale} shows how the median varies with bin size. All
three samples show larger amplitudes at longer binning. This is the
expected behaviour for red-noise light curves, in which longer
timescales carry more variance, and it is consistent with the power
spectral slopes measured in Section~\ref{sec:psd}. The CLB and FSRQ
curves remain close together throughout, and the separation from the
BL Lacs is preserved at every cadence.

\subsection{Matched controls}
\label{sec:matched}

Two properties of the CLB sample could in principle produce the observed
difference without any underlying difference in variability. We test
both at 30-day cadence, where the samples are largest.

The first is the distribution of synchrotron peak frequency.
\citet{Xiao2022} report that 90.7 per cent of CLBs are
low-synchrotron-peaked (LSP), whereas the BL Lac population spans all
three spectral classes. Since LSP and high-synchrotron-peaked \chg{(HSP)} sources
differ systematically in their variability, a comparison between samples
with different peak distributions could be misleading. Using \chg{the} synchrotron
peak frequencies from 4LAC-DR3 \citep{Ajello2022} and the boundaries of
\citet{Fan2016}, we classify every source \chg{as LSP,
intermediate-synchrotron-peaked (ISP) or HSP. The CLBs comprise 60 LSPs and 4 ISPs; the
BL Lacs 92 LSPs, 66 ISPs and 46 HSPs; and the FSRQs 386 LSPs and 3 ISPs.}
Restricting the comparison to LSP sources (Fig.~\ref{fig:ridgeline},
second row) does not weaken the result. The CLB--BL Lac probability becomes $3.0\times10^{-5}$, since LSP BL
Lacs are slightly less variable than the BL Lac sample as a whole, with
a median of 0.227 against 0.262. This likely reflects the
variability-index threshold of the repository, which admits faint
high-peaked sources only when they are strongly variable. The BL Lac control is therefore
not representative of the parent population in synchrotron peak, and we
draw no conclusions from the ordering of its spectral subclasses.

The second is brightness. \citet{Xiao2022} note that their sample is
brighter and more distant than \textit{Fermi} blazars in general, and
our CLBs inherit this property. Although $F_{\rm var}$ is
noise-debiased, the number of bins surviving the quality cuts depends on
flux, and this can leave a residual dependence. We therefore matched
each CLB to control sources in the plane of $\log$ mean flux against
$\log$ number of usable bins, taking nearest neighbours without
replacement and rejecting any match further than $0.25\sigma$ in the
standardised plane. Against FSRQs the matching succeeds: the flux
distributions agree with $p = 0.81$ and 81 per cent of CLBs find a
partner. Against BL Lacs it succeeds only partly ($p = 0.037$; 75 per
cent matched), because the two flux distributions overlap less well.

After matching, the CLB--BL Lac difference persists but is reduced, with
the KS statistic falling from $D = 0.308$ to $D = 0.259$
($p = 0.017$), while the CLB--FSRQ comparison remains consistent with no
difference ($p = 0.28$). Given the imperfect flux matching against
BL Lacs, some residual contribution from brightness to that comparison
remains possible, and we return to this in Section~\ref{sec:limits}.

\section{Spectral variability}
\label{sec:sigma}

The LCR provides a photon index in each time bin, so the same
noise-debiasing can be applied to the spectrum. We define the intrinsic
spectral variability as
\begin{equation}
\sigma_{\Gamma,\rm int}^{2} =
  {\rm Var}(\Gamma) - \overline{\sigma_{\Gamma}^{2}},
\label{eq:sigint}
\end{equation}
in direct analogy with equation~(\ref{eq:fvar}). This measures how much
the spectral shape of a source genuinely changes, as distinct from how
much the measured index scatters. No closed form corresponding to
equation~(\ref{eq:fvarerr}) is available for this quantity, so its
uncertainty is obtained by resampling the time bins of each source with
replacement. As in Table~\ref{tab:fvar}, values in brackets below are
the median per-source uncertainty.

\begin{figure*}
\centering
\includegraphics[width=\textwidth]{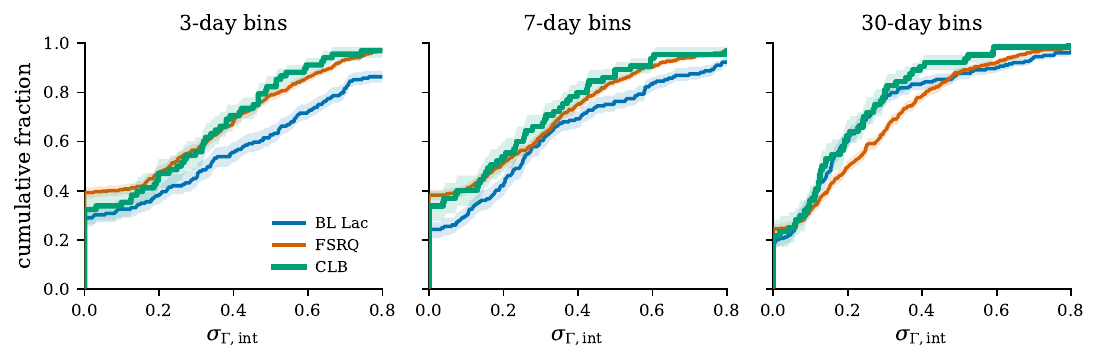}
\caption{Cumulative distributions of the noise-debiased spectral
variability $\sigma_{\Gamma,\rm int}$, at 3-day (left), 7-day (middle)
and 30-day (right) cadence. Colours, shaded bootstrap bands and legend
entries are as in Fig.~\ref{fig:fvar_ecdf}. Sources whose index scatter
is fully accounted for by measurement uncertainty return
$\sigma_{\Gamma,\rm int} = 0$ and appear as the step at the left-hand
edge of each panel; this affects the largest fraction of sources at
3-day cadence, where the per-bin index uncertainties are largest
relative to the intrinsic scatter, and the smallest fraction at 30-day
cadence. The three samples overlap closely at every cadence. No pair
differs significantly at 30- or 7-day cadence; at 3-day cadence CLBs
differ from BL Lacs at $p = 1.5 \times 10^{-2}$, in the sense that CLBs
are the less variable of the two.}
\label{fig:sigma}
\end{figure*}

At 30-day cadence the medians are 0.135 (0.058) for CLBs, 0.156 (0.055)
for BL Lacs and 0.202 (0.085) for FSRQs, and no pair of samples differs
significantly (Fig.~\ref{fig:sigma}). At 7-day cadence the medians are
0.167 (0.047), 0.246 (0.072) and 0.192 (0.071) and again no pair differs
significantly. At 3-day cadence they are 0.254 (0.055), 0.338 (0.070)
and 0.228 (0.068), respectively, and CLBs differ from BL Lacs at $p = 1.5 \times 10^{-2}$
, with CLBs showing lower spectral variability than BL Lacs. The nulls also hold against the SED- and
flux-matched controls of Section~\ref{sec:matched}. \chg{The spectral
variability of CLBs is therefore not significantly different from that of
either comparison sample,} and by
this measure the three populations behave alike. The ordering of the
three samples is not stable between cadences, so
$\sigma_{\Gamma,\rm int}$ does not place CLBs consistently with either
control. \chg{Stricter detection thresholds ($TS \ge 9$ or $TS \ge 25$)
do not change this: CLBs are never significantly more spectrally variable
than FSRQs, and the occasional differences from BL Lacs change sign
between cadences.}

One point of interpretation deserves comment. The \textit{raw} dispersion of the photon index differs between CLBs and FSRQs at all three cadences. For example, at 30-day cadence, the dispersions are 0.348 and 0.430 for CLBs and FSRQs, respectively. This difference tracks the corresponding median per-bin uncertainties of the two samples, \chg{0.25 and 0.29}, respectively, and disappears once the noise term is subtracted. It therefore reflects measurement precision rather than source behaviour, and we quote $\sigma_{\Gamma,\rm int}$ throughout.

\begin{figure}
\centering
\includegraphics[width=\columnwidth]{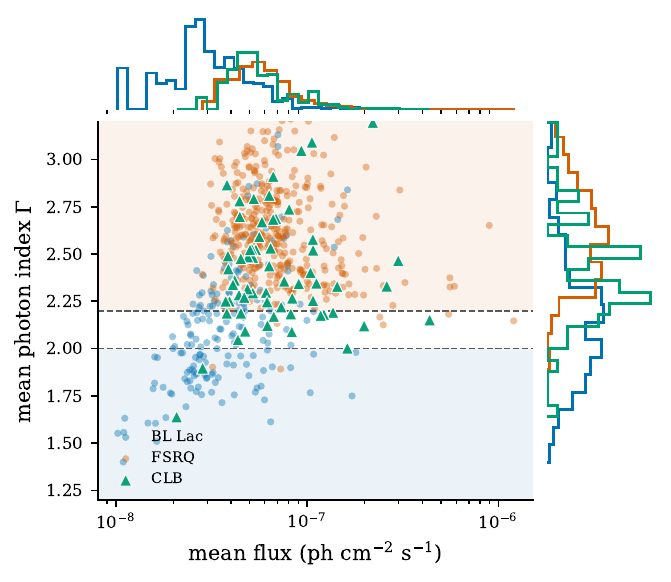}
\caption{Mean photon index against mean flux at 30-day cadence, for
sources with at least 20 bins carrying a well-measured photon index.
This is a slightly stricter requirement than the one used elsewhere in
the paper, which counts bins with a usable flux. Blue circles denote BL Lacs,
orange circles FSRQs and green triangles CLBs. Histograms above and to the right show the
marginal distributions of flux and of photon index for each sample,
normalised to unit area. The upper shaded band marks $\Gamma > 2.2$ and
the lower band $\Gamma < 2.0$, the values adopted by \citet{Ren2024b}
to identify FSRQ-like and BL Lac-like states; the unshaded strip between
them is the transition region. The two control samples separate in
photon index, with partial overlap in the transition strip, and overlap
substantially in flux.}
\label{fig:gammaflux}
\end{figure}

Fig.~\ref{fig:gammaflux} places the samples in the plane of mean photon
index against mean flux at 30-day cadence. The two control samples
separate along the vertical axis, as expected from their definition:
FSRQs lie mostly above $\Gamma = 2.2$ and BL Lacs mostly below it,
many of them in the strip between 2.0 and 2.2. These are the boundaries
used by \citet{Ren2024b} to identify spectral states. The CLBs fall predominantly in the upper band, among
the FSRQs, rather than occupying the transition strip between the two.
Their mean photon index therefore does not sit at an intermediate value,
which is the first indication that their $\gamma$-ray spectra are
FSRQ-like rather than mixed. The horizontal axis shows that the three
samples overlap substantially in flux, although the CLBs are on average
somewhat brighter, which motivates the matched comparison of
Section~\ref{sec:matched}.

\section{Flux and index distributions, and power spectra}
\label{sec:dist}

\subsection{Distribution models}
\label{sec:models}

The shape of a flux distribution carries information about the process
generating the variability. A normal distribution arises when many
independent additive perturbations combine, whereas a lognormal
distribution indicates that the perturbations act multiplicatively, as
expected if fluctuations propagate inwards through an accretion flow
\citep{Lyubarskii1997, Uttley2005, Scaringi2012}. Lognormal flux
distributions have been reported for blazars in $\gamma$-rays
\citep{Shah2018, Shah2020, Sinha2018} and in X-rays
\citep{Khatoon2020}. A distribution requiring two lognormal components
suggests the presence of two distinct emission states. This is of
obvious interest for sources thought to switch between states. Such
double-lognormal profiles have been reported in bright \textit{Fermi}
blazars \citep{Shah2018, Shah2025}, and in both flux and photon index
in very-high-energy FSRQs \citep{Malik2025}.

We fit three models to each distribution by unbinned maximum likelihood:
a normal distribution, a single lognormal, and a two-component lognormal
mixture fitted by expectation maximisation. Because all three are
densities over the same variable, their Bayesian information criteria
(BIC) can be compared directly, and we adopt the model with the lowest
BIC. Fitting without binning avoids any dependence of the result on the
choice of histogram bins.

A source is fitted if at least 30 of its bins survive the quality
cuts. The comparison is made at all three cadences. This matters because the mixture carries three more free parameters than
the single lognormal. At the same time the number of bins per source
falls as the bins are shortened, so the model choice is best determined
at 30-day cadence and progressively noisier at 7- and 3-day. We therefore treat the 30-day result as
primary and read the shorter cadences alongside it.

\begin{figure}
\centering
\includegraphics[width=\columnwidth]{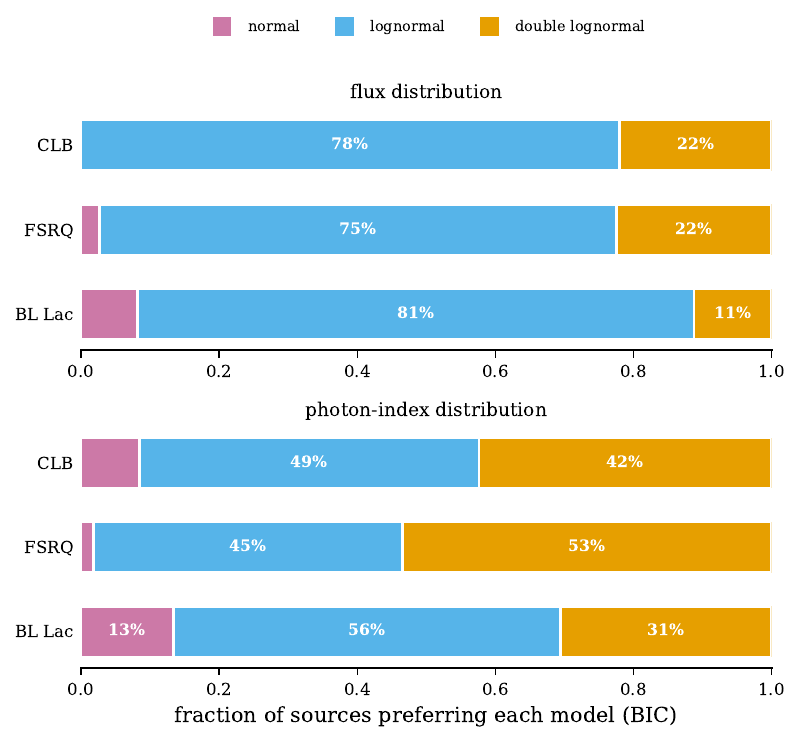}
\caption{Fraction of sources in each sample for which each of the three
models gives the lowest BIC, at 30-day cadence, for the flux
distribution (upper panel) and the photon-index distribution (lower
panel). Pink denotes a normal distribution, blue a single lognormal and
orange a two-component lognormal mixture. Percentages are printed within
each segment where there is room. The flux distributions are predominantly
lognormal in all three samples, whereas the index distributions are only
rarely normal.}
\label{fig:models}
\end{figure}

\begin{figure*}
\centering
\includegraphics[width=\textwidth]{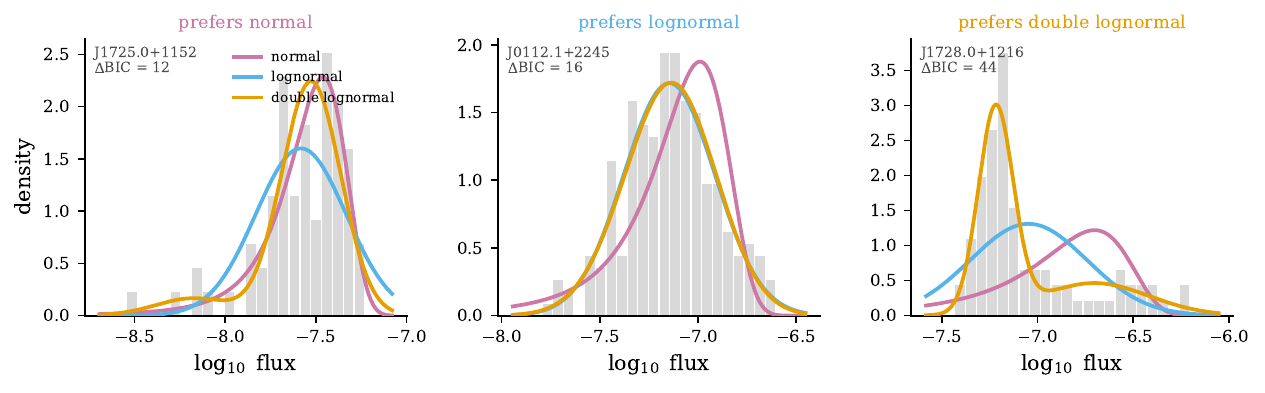}
\caption{Three representative flux distributions at 30-day cadence, one
preferring each model, shown as grey histograms of $\log_{10}$ flux.
\chg{For each model the example is the source, from any of the three
samples, for which the BIC preference is strongest; they are the BL Lacs
4FGL~J1725.0$+$1152 (normal) and 4FGL~J0112.1$+$2245 (lognormal) and the
FSRQ 4FGL~J1728.0$+$1216 (two-component mixture), so none is a CLB. The
figure illustrates what each preference looks like in the data rather
than the behaviour of any one sample.}
All three fitted models are overlaid in each panel for comparison: pink
for the normal distribution, blue for the single lognormal and orange
for the two-component mixture. The panel title states which model the
BIC selects, and the source name and the BIC difference to the
second-best model are given at upper left. A large BIC difference, as in
the right-hand panel, corresponds to a visibly two-peaked distribution
rather than to a marginal preference.}
\label{fig:examplefits}
\end{figure*}

For the flux distributions at 30-day cadence, a single lognormal is
preferred for most sources in every sample, in agreement with earlier
work on smaller samples \citep{Shah2018, Shah2020}. A two-component
mixture is preferred for 22.0 per cent of the 59 CLBs, 22.5 per cent of
the 338 FSRQs and 11.3 per cent of the 160 BL Lacs with enough usable
bins to be fitted (Fig.~\ref{fig:models}, upper panel).
Fig.~\ref{fig:examplefits} shows one example of each case, so that the
strength of the preference can be judged directly.

The mixture fraction rises at shorter cadence in every sample, as
expected when longer bins average over shorter-timescale structure. It
reaches 36.7 per cent of CLBs, 41.1 per cent of FSRQs and 25.4 per cent
of BL Lacs at 7-day cadence, and 41.0, 42.6 and 29.1 per cent at 3-day.
This is the behaviour \citet{Malik2025} report for very-high-energy
FSRQs, where double-lognormal profiles are clearest at 3- and 7-day
binning.

At no cadence do CLBs show an excess of bimodal flux distributions over
FSRQs. A one-sided Fisher exact test gives $p = 0.59$, $0.79$ and
$0.65$ at 30-, 7- and 3-day cadence, and at all three cadences the
CLB fraction is in fact the lower of the two. Against BL Lacs the CLB
fraction is the higher of the two, significantly so at 30-day cadence
($p = 0.038$) but not at 7- or 3-day ($p = 0.080$ and $0.074$); that
ordering is the one expected if CLBs behave as FSRQs rather than as BL
Lacs. Bimodality might be expected for sources alternating between two
emission states, so the comparison against the FSRQ control is the
informative one here, and it returns no excess on any timescale we can
probe.

\chg{Gaps in a light curve could in principle mimic bimodality, if they
remove bins while the flux is steadily rising or falling and so leave two
separated groups of flux values. We tested for this in three ways. First,
the CLB light curves do not have more gaps than those of the controls: at
30-day cadence the median fraction of time bins without a usable flux is
0.51 for CLBs, against 0.68 for FSRQs and 0.64 for BL Lacs. Second, we
imposed the CLB sampling on the controls. All LCR light curves share
a common time grid at a given cadence, so the sampling pattern of any CLB
can be imposed on any control source by keeping only the bins at which
that CLB has a usable flux. For each control source we did this for five
randomly chosen CLBs and refitted the three models. At 30-day cadence the
fraction of FSRQ light curves preferring the mixture changes from 22.5 to
21.3 per cent once CLB gaps are imposed, and that of BL Lac light curves
from 11.3 to 7.6 per cent; at 7- and 3-day cadence the fractions fall by
4--12 percentage points. Imposing CLB gaps therefore never increases the
preference for two components, and usually reduces it, as expected when
fewer bins weaken the evidence for the extra parameters. Third, within
the CLB sample, the sources that prefer the mixture have, if anything,
fewer gaps than the rest (median fraction of bins missing 0.46 against
0.51 at 30-day cadence). Gaps are therefore not responsible for the mixture
fractions reported above.}

For the photon-index distributions, \chg{fitted with the same
requirement of at least 30 usable bins so that the flux and index
fractions rest on comparable samples,} a normal distribution is preferred
for only 2--13 per cent of sources, so index distributions are generally
not Gaussian. This agrees with the behaviour reported by
\citet{Shah2025} and \citet{Malik2025} for other blazar samples, and
contrasts with \chg{the X-ray behaviour of 3C~273, whose index distribution
is close to Gaussian while its flux distribution is lognormal
\citep{Khatoon2020}. A likely reason for the difference is measurement
precision. The per-bin uncertainty on the $\gamma$-ray index (median 0.25
at 30-day cadence) is larger than the intrinsic index scatter of most
sources (Section~\ref{sec:sigma}) and varies strongly from bin to bin with
$TS$, so the observed distribution is the intrinsic one broadened by
errors of very different sizes. Such a superposition has heavier tails
than a single Gaussian even when the intrinsic distribution is Gaussian,
and is therefore better described by the more flexible models. The X-ray
indices of 3C~273, a bright source monitored with \textit{RXTE}, are
measured with much greater precision relative to their intrinsic
scatter.}
Table~\ref{tab:indexmodels} gives the fractions at all three cadences.
\chg{However, two caveats apply to this analysis. First, over} the narrow range that $\Gamma$ spans, a
lognormal \chg{distribution} is numerically close to a \chg{normal distribution}, so the two are not always
well separated. \chg{Second, the} preferred model for a given source
agrees between cadences in only 54--68 per cent of cases. The
distinction between normal and lognormal should therefore not be
over-interpreted, although the preference for two components over one is
more robust.

\begin{table}
\centering
\caption{Fraction of sources for which each model gives the lowest BIC
for the photon-index distribution, at each cadence. $N$ is the number of
sources with enough usable bins to be fitted.}
\label{tab:indexmodels}
\begin{tabular}{llccccc}
\hline
Cadence & Sample & $N$ & Normal & Lognormal & Double\\
\hline
30\,d & BL Lac & 157 & 0.134 & 0.561 & 0.306\\
      & FSRQ   & 335 & 0.018 & 0.448 & 0.534\\
      & CLB    &  59 & 0.085 & 0.492 & 0.424\\
\hline
7\,d  & BL Lac & 125 & 0.112 & 0.400 & 0.488\\
      & FSRQ   & 366 & 0.066 & 0.462 & 0.473\\
      & CLB    &  58 & 0.069 & 0.397 & 0.534\\
\hline
3\,d  & BL Lac & 127 & 0.110 & 0.378 & 0.512\\
      & FSRQ   & 395 & 0.078 & 0.446 & 0.476\\
      & CLB    &  61 & 0.082 & 0.393 & 0.525\\
\hline
\end{tabular}
\end{table}

\subsection{Centroid and width}
\label{sec:musigma}

Section~\ref{sec:models} asked which model best describes each
distribution. A second and separate question is how the distributions of
different sources relate to one another: whether the sources that are
brighter on average are also the ones that vary most, and likewise
whether the sources with softer mean spectra are the ones whose spectra
vary most. A source whose variability amplitude scales with its mean
flux behaves multiplicatively in the sense described above, and a
population in which the two quantities are unrelated does not.

\begin{figure*}
\centering
\includegraphics[width=\textwidth]{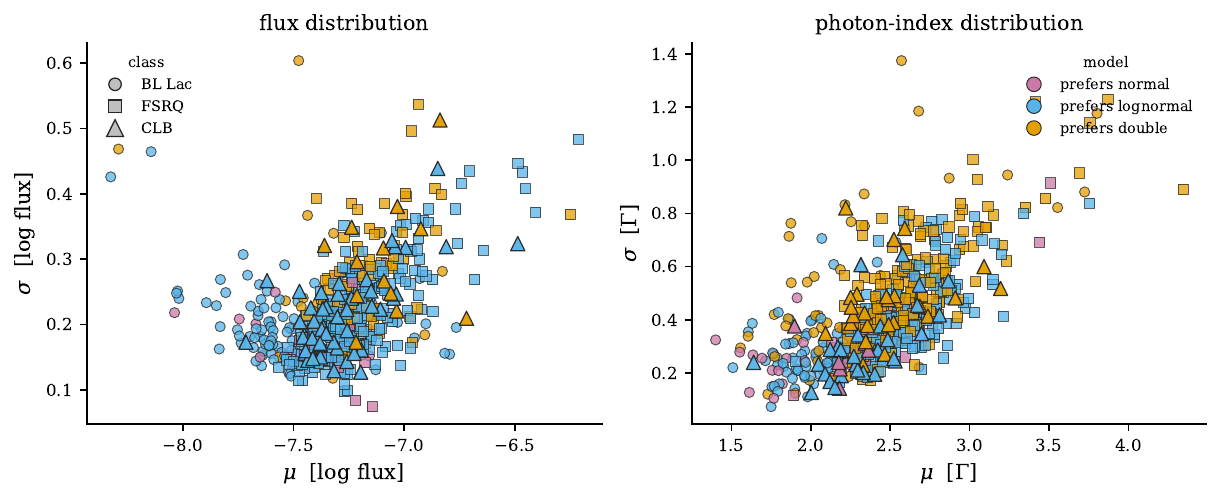}
\caption{Centroid $\mu$ against width $\sigma$ of a single lognormal
fitted to the flux distribution (left) and to the photon-index
distribution (right), at 30-day cadence. Marker shape denotes the
sample, with circles for BL Lacs, squares for FSRQs and triangles for
CLBs. Marker colour denotes the model that the BIC actually prefers for
that source, using the same colours as Fig.~\ref{fig:models}: pink for
normal, blue for lognormal and orange for a two-component mixture.}
\label{fig:musigma}
\end{figure*}

To ask this we need a single pair of numbers per source. We therefore
fit a single lognormal to every source and use its centroid $\mu$ and
width $\sigma$, so that the two sets of measurements are comparable. This model is fitted
to all sources, including those for which the BIC prefers a mixture,
because the purpose here is to place every source on a common plane
rather than to identify the best description of each one. The cost is
that a single lognormal fitted to a genuinely two-peaked distribution
returns an artificially large width, since one component is stretched
across both peaks. Such sources are identified by colour in
Fig.~\ref{fig:musigma}, so that their contribution to the plane can be
seen rather than assumed to be negligible.

The relation between $\mu$ and $\sigma$ is shown in
Fig.~\ref{fig:musigma}. For the flux distributions at 30-day cadence
the correlation is strong for FSRQs ($\rho = +0.581$,
$p = 5.9\times10^{-32}$) and consistent with zero for BL Lacs
($\rho = -0.034$, $p = 0.67$). CLBs follow the FSRQs
($\rho = +0.557$, $p = 4.6\times10^{-6}$). The same pattern holds at
the other two cadences: for BL Lacs $\rho = +0.057$ and $+0.030$ at 7-
and 3-day cadence, neither significant, against $+0.562$ and $+0.467$
for FSRQs and $+0.592$ and $+0.456$ for CLBs.

Two conclusions follow. First, in FSRQs and CLBs the brighter sources
are also the more variable ones, in the sense that the width of the flux
distribution grows with its centroid, whereas in BL Lacs the two are
unrelated. This difference between the control samples is itself
consistent with the multiplicative picture of
Section~\ref{sec:models}. Nor is it an artefact of the sources with
two-component distributions: these are a minority in each sample, and
are distributed through the plane rather than concentrated at one end of
it. Second, and for our purposes more importantly, CLBs are
indistinguishable from FSRQs in this respect and clearly separated from
BL Lacs, which adds a further diagnostic to those of
Sections~\ref{sec:fvar} and~\ref{sec:sigma}.

The index distributions behave differently. There all three samples show
a positive correlation of similar strength, with $\rho$ between $+0.45$
and $+0.65$ at 30-day cadence, so this plane does not distinguish
between the populations and we do not use it further.

\subsection{Power spectral density}
\label{sec:psd}

The power spectral density describes how the variance of a light curve
is distributed across timescales, and so complements $F_{\rm var}$,
which measures only the total. Power spectra were computed using
Bartlett's method, dividing each light curve into equal segments,
computing a periodogram for each and averaging them, which reduces the
scatter inherent in a single periodogram at the cost of frequency
resolution.

We fit a simple power law, $P(f) \propto f^{-\beta}$, to the averaged
periodogram. This form is used because blazar power spectra are
generally well described by a featureless red-noise power law over the
frequency range accessible to \textit{Fermi}, with no evidence for a
characteristic timescale that would appear as a break
\citep{Abdo2010, Sobolewska2014, Meyer2019}. The slope $\beta$ then
summarises the balance between long and short timescales: a larger
$\beta$ means that a greater share of the variance is carried by slow
variations, while $\beta \simeq 0$ corresponds to white noise.

\begin{figure}
\centering
\includegraphics[width=\columnwidth]{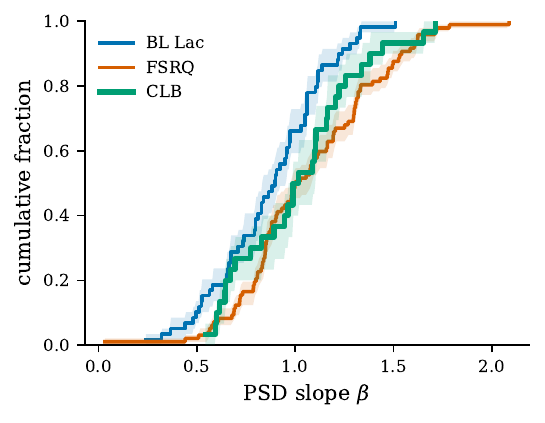}
\caption{Cumulative distributions of the fitted power spectral density
slope $\beta$ at 30-day cadence, where $P(f) \propto f^{-\beta}$. Blue
denotes BL Lacs, orange FSRQs and green CLBs, and the number of sources
contributing to each curve is given in Section~\ref{sec:psd}; only sources with
enough usable bins to sustain a segmented periodogram are included.
The shaded band around each curve is the 68 per cent confidence region
from bootstrap resampling of the sources. Larger $\beta$ means a greater
share of the variance is carried by slow variations, while
$\beta \simeq 0$ would correspond to white noise. All three
distributions lie in the range expected for red-noise light curves and
overlap one another, and no pair differs significantly.}
\label{fig:psd}
\end{figure}

At 30-day cadence the median slope is $1.003$ for the 30 CLBs with
sufficient sampling, $0.898$ for 59 BL Lacs and $1.001$ for 97 FSRQs
(Fig.~\ref{fig:psd}). No pair of samples differs significantly
($p = 0.14$ for CLB against BL Lac and $p = 0.59$ for CLB against FSRQ),
and the CLB value lies close to the FSRQ one. These slopes are
consistent with those reported for blazars over comparable baselines
\citep{Sobolewska2014, Meyer2019, Shah2025}, and their being close to
unity is consistent with the increase of $F_{\rm var}$ with bin size
seen in Fig.~\ref{fig:timescale}. We restrict this measurement to 30-day
cadence because a reliable power-law fit requires a long, well-sampled
series: after the quality cuts, only four CLBs at 3-day cadence and six
at 7-day cadence satisfy the length requirement, which is too few for a
meaningful comparison. \chg{By this measure, therefore, CLBs are
indistinguishable from both FSRQs and BL Lacs.}

\section{The photon-index/flux relation}
\label{sec:rho}
\label{sec:indexflux}

The way in which the spectrum responds to a change in flux is a more
direct probe of the emission mechanism than the amplitude of the
variability alone. A spectrum that softens as the source brightens
suggests that radiative cooling keeps pace with particle injection,
whereas hardening suggests efficient acceleration or a growing
external-Compton component \citep{KirkRieger1998, Sikora1994}. Both
behaviours are seen in blazars, sometimes in the same source at
different epochs \citep{Shah2019, Akbar2024, Malik2025}. For each source
we therefore compute the Spearman correlation between photon index and
flux across its light curve.

\begin{figure*}
\centering
\includegraphics[width=\textwidth]{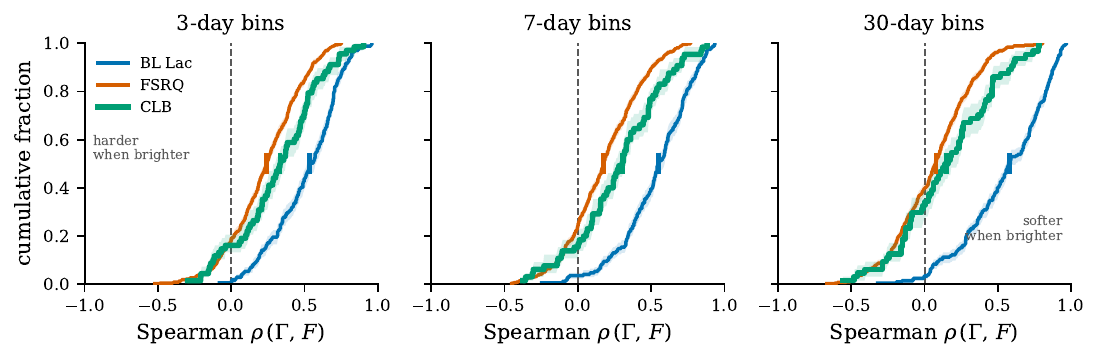}
\caption{Cumulative distributions of the per-source Spearman correlation
$\rho$ between photon index and flux, at 3-day (left), 7-day (middle)
and 30-day (right) cadence. Colours, shaded bootstrap bands, median
ticks and legend entries are as in Fig.~\ref{fig:fvar_ecdf}. The dashed
vertical line marks $\rho = 0$: sources to its right soften as they
brighten and those to its left harden. The CLB curve lies between the
two control curves at all three cadences, closer to the FSRQ curve than
to the BL Lac curve.}
\label{fig:rho}
\end{figure*}

At 30-day cadence the median per-source correlation is $+0.580$ for
BL Lacs, $+0.077$ for FSRQs and $+0.149$ for CLBs
(Fig.~\ref{fig:rho}, Table~\ref{tab:rho}). BL Lacs thus show a clear
softer-when-brighter behaviour while FSRQs show only a weak one, in
agreement with previous work \citep{Shah2025}. CLBs lie between the two.
They differ from BL Lacs at $p = 3.7 \times 10^{-12}$ and from FSRQs at
$p = 0.042$, and the same pattern, with the same sign, is found at 7-
and 3-day cadence, where the CLB medians are $+0.304$ and $+0.334$. The
difference from FSRQs persists after flux and sampling matching
($+0.149$ against $+0.080$, $p = 0.026$), in the comparison where the
matching succeeds, so it is unlikely to follow from the greater
brightness of the CLB sample. \chg{It is, however, sensitive to the
detection threshold: it remains with $TS \ge 25$ imposed on every bin
($p = 0.038$) but is not significant with $TS \ge 9$ ($p = 0.22$), so we
regard it as suggestive.}

\begin{table}
\centering
\caption{Median per-source Spearman correlation between photon index and
flux, and the population-level correlation between mean index and mean
flux, at 30-day cadence. \chg{Below the per-source values, $p$ is the KS
probability that the per-source correlations of CLBs and of that control
share a parent population; below the population-level values, $p$ is the
probability of the correlation arising by chance.} The last two rows give
the population-level values reported by \citet{Shah2025} for the same
quantity and cadence.}
\label{tab:rho}
\begin{tabular}{lccc}
\hline
 & CLB & FSRQ & BL Lac\\
\hline
Per-source $\rho$ & $+0.149$ & $+0.077$ & $+0.580$\\
\chg{\quad $p$ (KS, against CLB)} & -- & \chg{0.042} & \chg{$3.7\times10^{-12}$}\\
Population $\rho$ & $+0.011$ & $-0.175$ & $+0.485$\\
\chg{\quad $p$ (correlation)} & \chg{0.93} & \chg{$5.1\times10^{-4}$} & \chg{$2.0\times10^{-13}$}\\
\citet{Shah2025}, population $\rho$ & -- & $-0.12$ & $+0.47$\\
\chg{\quad $p$ (correlation)} & -- & \chg{$4.9\times10^{-3}$} & \chg{$4.3\times10^{-23}$}\\
\hline
\end{tabular}
\end{table}

The population-level relation, obtained by correlating each source's
mean index with its mean flux, gives a different perspective. Here CLBs
behave as FSRQs do, with $\rho = +0.011$ ($p = 0.93$) against $-0.175$
for FSRQs and $+0.485$ for BL Lacs. Our control values are close to
those reported by \citet{Shah2025} for the same quantity, which supports
the measurement.

Taken together, the two levels give a consistent picture. Across
sources, CLBs sit where FSRQs sit in the index--flux plane; within a
source, their spectra respond to brightening slightly more than FSRQ
spectra do, though much less than BL Lac spectra. The per-source
correlation is therefore the one time-averaged diagnostic on which CLBs
are displaced from the FSRQ sample, and we return to it in
Section~\ref{sec:interp}.

\section{Time-resolved behaviour}
\label{sec:time}

The measurements presented so far summarise each light curve by a single
number. We now examine the light curves as functions of time. All
figures in this section use the 30-day light curves, since these give
the best-determined photon index per bin, which is the quantity of
interest here.

\begin{figure*}
\centering
\includegraphics[width=\textwidth]{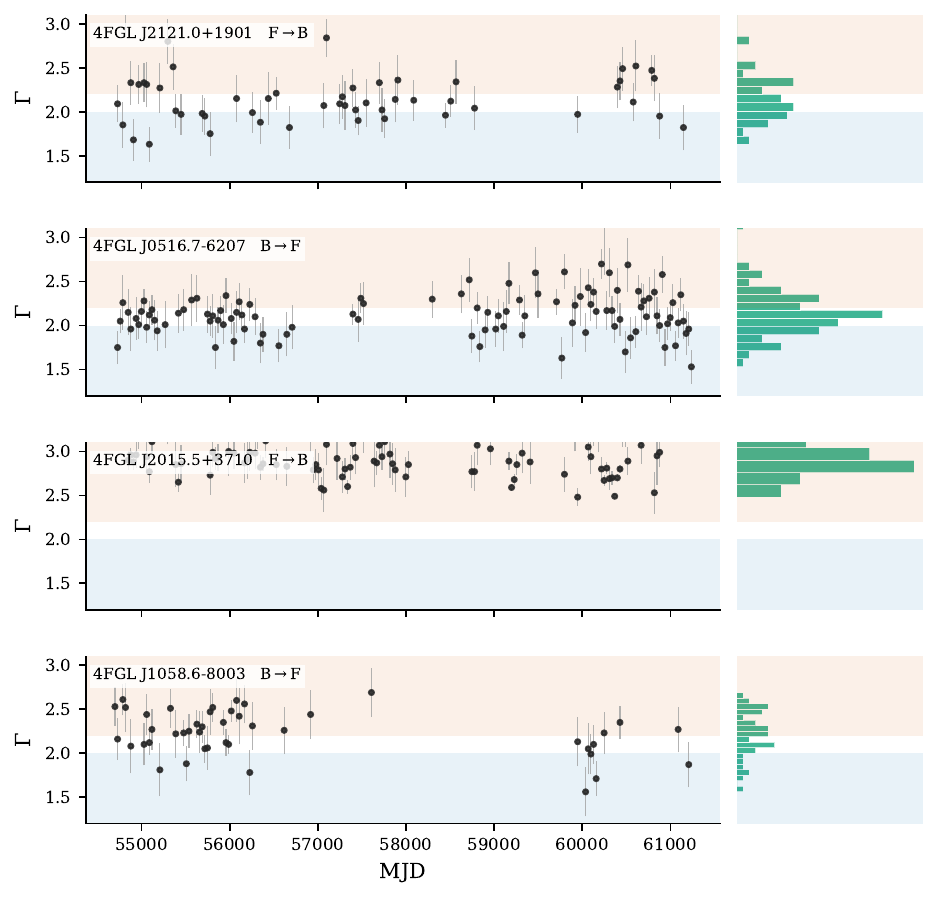}
\caption{Photon index against time, at 30-day cadence, for the four
tier~A CLBs \chg{which have} the largest spectral excursions, measured as the range
between the 5th and 95th percentiles of $\Gamma$. Points show individual
time bins with their $1\sigma$ uncertainties; only bins passing the
quality cuts and with $TS \ge 25$ are plotted. The upper shaded band
marks $\Gamma > 2.2$ and the lower $\Gamma < 2.0$, as in
Fig.~\ref{fig:gammaflux}. The source name and, where recorded, the
direction of its optical transition are given at upper left. The panel \chg{on the
right} of each light curve is the marginal distribution of $\Gamma$ for
that source, drawn on the same vertical scale, so that the shape of the
distribution can be compared directly with the position of the state
bands. The four sources shown are, from top to bottom,
4FGL~J2121.0$+$1901, 4FGL~J0516.7$-$6207, 4FGL~J2015.5$+$3710 and
4FGL~J1058.6$-$8003. All four marginal distributions are singly peaked.
The third source lies entirely within the FSRQ-like band, and the gap
between MJD~57000 and 59900 in the fourth illustrates the uneven
sampling of the fainter sources.}
\label{fig:lightcurves}
\end{figure*}

\begin{figure*}
\centering
\includegraphics[width=\textwidth,height=0.78\textheight,keepaspectratio]{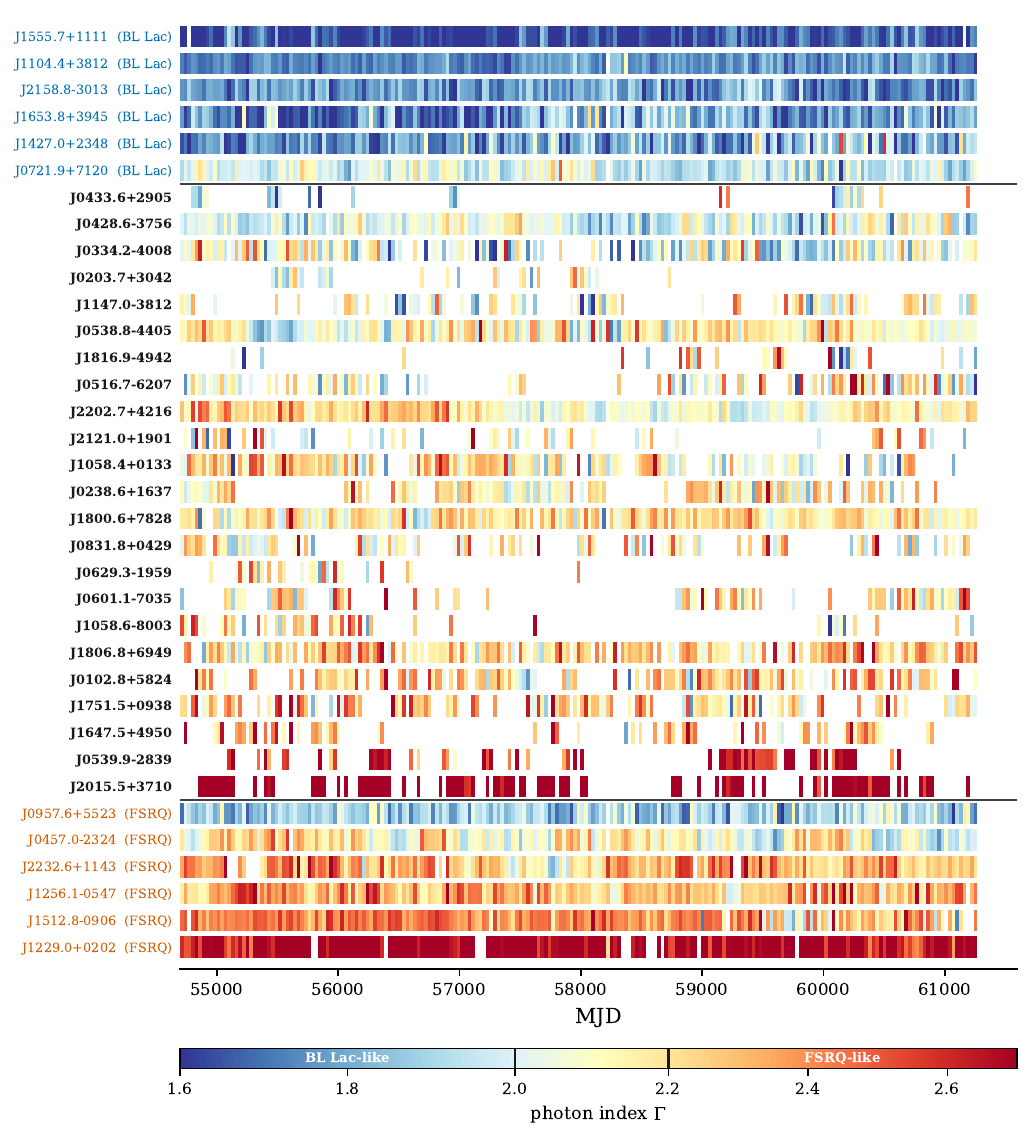}
\caption{Photon index as a function of time for the 23 tier~A CLBs with
at least 25 time bins satisfying $TS \ge 25$, with one row per source and
one coloured cell per 30-day time bin. This threshold is stricter than
that used elsewhere in the paper, and is imposed so that each row
contains enough well-measured cells to be read; the remaining tier~A
sources are too sparsely sampled to display in this form. Colour
encodes the measured photon index on the diverging scale shown at the
bottom, which is centred on $\Gamma = 2.1$, the middle of the transition
region; blue corresponds to hard, BL Lac-like spectra and red to soft,
FSRQ-like spectra. Vertical black lines on the colour bar mark
$\Gamma = 2.0$ and $\Gamma = 2.2$. Gaps indicate bins that fail the
quality cuts. CLB source names are shown in bold; above and below the
horizontal separators are the best-sampled BL Lac and FSRQ control
sources, labelled and coloured by sample, included so that CLB
behaviour can be judged against sources of known and stable spectral
type. Rows are ordered by mean photon index within each group, so that the
progression from hard to soft spectra can be followed down the panel.
The control rows provide a visual reference: a source of stable spectral
type appears as a row of nearly uniform colour, so departures from
uniformity in the CLB rows can be judged against sources of known
classification rather than in isolation. Control sources can depart from
the colour typical of their class: 4FGL~J1229.0$+$0202 (3C~273) is
uniformly soft, while 4FGL~J0957.6$+$5523 (4C~$+$55.17) appears
predominantly hard and has had its FSRQ classification questioned
\citep{McConville2011}.}
\label{fig:timeline}
\end{figure*}

Fig.~\ref{fig:lightcurves} shows the photon index against time for the
four tier~A CLBs \chg{which have} the largest spectral excursions, together with the
distribution of $\Gamma$ for each. These are the sources in which a
change of spectral state, if present, should be easiest to see. The
marginal distributions are informative in this respect. A source
alternating between two states would produce a two-peaked distribution
with the peaks falling in the two shaded bands, whereas a source varying
about a single spectral type produces one peak.

None of the four shows two separated peaks. In
4FGL~J2121.0$+$1901 and 4FGL~J1058.6$-$8003 the index ranges from about
1.5 to 2.8 and the marginal distribution is broad and singly peaked,
straddling the transition region rather than avoiding it. In
4FGL~J0516.7$-$6207 the distribution is likewise single-peaked, with a
maximum just above $\Gamma = 2.2$. The fourth source,
4FGL~J2015.5$+$3710, is the most striking. Every one of its measured
bins lies above $\Gamma \simeq 2.4$, so the source remains within the
FSRQ-like band throughout the mission and never enters the BL Lac-like
band at all, despite being recorded as an $F\!\rightarrow\!B$ transition
in the optical. Its $\gamma$-ray spectrum gives no indication of the
change of look inferred from its optical spectra.

The sampling also varies considerably between sources.
4FGL~J1058.6$-$8003 has almost no usable bins between MJD~57000 and
59900, so a state change occurring in that interval would not be
recorded. This is a general limitation of fixed-cadence light curves for
faint sources and is one reason why we do not attempt to identify
individual transitions.

Fig.~\ref{fig:timeline} presents the better-sampled tier~A sources as
strips coloured by photon index, with six BL Lac and six FSRQ control
sources included for comparison. Displaying a light curve in this form
requires enough well-measured bins for a row to be legible, and we
therefore restrict the figure to the 23 tier~A sources with at least 25
bins satisfying $TS \ge 25$. The figure serves two purposes. It shows
the full time history of each of these sources in a single panel, so
that a sustained change of spectral state would appear as an extended
block of contrasting colour rather than as isolated bins. It also makes the
sampling explicit: gaps mark bins failing the quality cuts, and these
are frequent for the fainter sources.

\chg{In Fig.~\ref{fig:timeline} each cell is coloured by the photon index
of that 30-day bin on a diverging scale centred on $\Gamma = 2.1$, so that
blue cells are hard (BL Lac-like, $\Gamma < 2.0$), red cells soft
(FSRQ-like, $\Gamma > 2.2$) and pale cells lie in the transition region
between them; blank cells are bins without a usable measurement.} Most CLB rows are predominantly warm in colour, and in this respect
resemble the FSRQ control rows more closely than the BL Lac ones.
Individual blue bins occur throughout, but extended blue blocks are
uncommon, and no row shows the clean two-block structure that a
long-lived change of state would produce. The rows are ordered by mean
photon index, and the progression from the hardest CLBs at the top to
4FGL~J2015.5$+$3710 at the bottom is gradual rather than divided into
two groups.

\chg{The control rows are instructive in their own right,} because they show how far the
$\gamma$-ray photon index can vary within a single optical class. Among
the BL Lacs, 4FGL~J1555.7$+$1111 is uniformly hard, whereas
4FGL~J0721.9$+$7120 sits close to the transition region for most of the
mission. Among the FSRQs the range is wider still. 4FGL~J1229.0$+$0202
(3C~273) is uniformly soft, as expected, but 4FGL~J0957.6$+$5523
\chg{(4C~$+$55.17)} appears predominantly hard and would be classified as BL Lac-like on
photon index alone. \chg{This source} is well known for
having an unusually hard and remarkably steady $\gamma$-ray spectrum for
an FSRQ; its classification has itself been questioned, and
\citet{McConville2011} considered whether it might instead be a young
radio source \citep[see also][]{Ajello2020}.

This is a useful reminder rather than an inconsistency. The 4FGL class
is assigned from optical spectroscopy and does not fix the $\gamma$-ray
photon index, so a criterion based on $\Gamma_{\gamma}$ alone will not
reproduce the optical classification for every source. It reinforces the
caution set out below in Section~\ref{sec:occupancy} about applying such a
criterion bin by bin across a population.

\subsection{Directional test}
\label{sec:directional}

Tier~A sources carry a transition direction assigned from optical
spectroscopy without reference to $\gamma$-ray data. This permits a
prediction that can be tested independently. If the optical transition
reflects a change in the jet, then $B\!\rightarrow\!F$ sources should
soften in $\gamma$-rays over the mission and $F\!\rightarrow\!B$ sources
should harden.

We fitted an inverse-variance weighted linear trend to $\Gamma(t)$ for
each source. \chg{For the tier~A sources with sufficient sampling at 30-day
cadence, $\approx 60$\,per cent (17 of 28 sources) have a slope of the
predicted sign ($p = 0.17$ from a one-sided binomial test); at 3-day
cadence this falls to 50 per cent (11 of 22 sources, $p = 0.58$).}
The optical transition direction therefore provides no statistically
significant prediction of the sense of $\gamma$-ray spectral evolution.

Two checks indicate that this is not a failure of sensitivity. The
control samples, which carry no direction, have slope distributions
symmetric about zero, with 52.5 \chg{(BL Lacs)} and 48.4 per cent \chg{(FSRQs)} positive slopes,
providing no evidence for a systematic sign bias in the slope measurements.
Moreover several individual sources show highly significant secular
evolution; BL~Lacertae itself has a slope significant at $15\sigma$. The
$\gamma$-ray spectra of these objects do evolve, but not in step with
their optical classification. We note one limitation: the catalogue
records the direction of each transition but not its epoch, so the test
uses a trend over the full mission rather than a comparison across the
transition itself, which reduces its sensitivity to a change confined to
a short interval. \chg{The optical epochs are also widely separated:
the tier~A equivalent widths come from single spectra published between
1983 and 2020 \citep{Paliya2021}, so a change of classification may have
taken place at any time over years or decades, possibly before the LAT
mission began.}

\subsection{State occupancy}
\label{sec:occupancy}

\citet{Ren2024b} proposed that the $\gamma$-ray photon index can be used
to identify the state of a changing-look source, with
$\Gamma_{\gamma} \lesssim 2.0$ corresponding to a BL Lac-like state,
$\Gamma_{\gamma} \gtrsim 2.2$ to an FSRQ-like state and the interval
between to a transitional state. The criterion was developed for
OQ~334. Its lower boundary comes from comparing the indices of
\citet{Ren2024a}, measured over epochs defined by the Mg\,\textsc{ii}
equivalent widths of \citet{Mishra2021}, with those widths. Its upper
boundary follows the population division of \citet{2010ApJ...710.1271A}. The
epochs of \citet{Ren2024b} postdate the last of those spectra and are
defined from the $\gamma$-ray light curve.

Applying the criterion to individual light-curve bins is less
straightforward. The epochs of \citet{Ren2024b} span 4 to 767 days,
with a median near 40 days, and the index uncertainty in all but one
lies between 0.02 and 0.15. Our per-bin uncertainties at 30-day cadence
have a median of $\sigma_{\Gamma} = \chg{0.25}$ for the CLBs, which exceeds
the 0.2-wide transition interval, so an individual bin cannot be
assigned to a state with confidence. Simulations using the empirical
distribution of $(TS, \sigma_{\Gamma})$ confirm that simply counting
boundary crossings is dominated by measurement noise. The two
approaches are therefore complementary rather than competing: theirs
achieves precision by integrating over variable-length epochs for one
well-observed source, ours achieves uniform coverage of a population at
fixed cadence.

We therefore adopt a conservative criterion, requiring a bin to lie
$2\sigma$ clear of a boundary and a state to be counted as occupied only
if at least two such bins fall within it. On this basis 14.0 per cent of
CLBs occupy both states, against 1.6 per cent of BL Lacs (Fisher exact
test, $p = 0.0034$) and 6.7 per cent of FSRQs ($p = 0.096$). The result
survives only four of nine combinations of $TS$ threshold and
significance level, so we regard it as suggestive rather than
established. A population-level application of the \citet{Ren2024b}
criterion would need optically defined epochs for each source, which are
presently available for very few objects.

\begin{figure}
\centering
\includegraphics[width=\columnwidth]{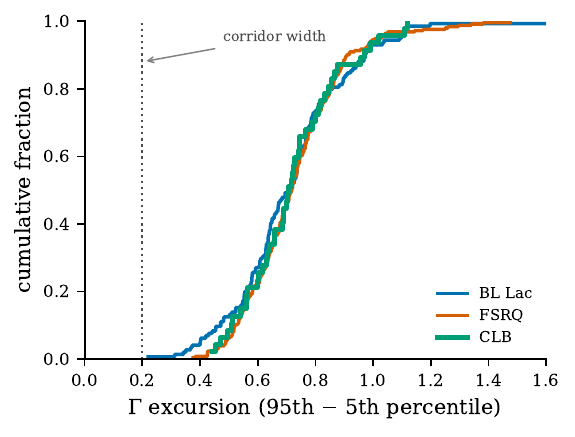}
\caption{Cumulative distributions of the range in photon index traversed
by each source at 30-day cadence, measured between the 5th and 95th
percentiles of $\Gamma$ so that a single discrepant bin cannot inflate
the value. Colours are as in Fig.~\ref{fig:fvar_ecdf}. The excursion is measured
only for sources having at least 20 bins that carry a well-measured
photon index and at least 20 bins satisfying $TS \ge 25$. The dotted vertical line marks 0.2, the
width of the transition interval between $\Gamma = 2.0$ and
$\Gamma = 2.2$. Sources to the left of this line cannot cross the
interval at all; the large fraction of sources lying well to its right
shows that a wide excursion is common in every sample and is therefore
not by itself a signature of changing-look behaviour.}
\label{fig:excursion}
\end{figure}

A related quantity is the range in photon index that each source
actually covers, shown in Fig.~\ref{fig:excursion}. In every sample the
majority of sources traverse a range comparable to or larger than the
width of the transition interval. Wide excursions in $\Gamma$ are
therefore common among ordinary blazars, and their presence in a CLB
light curve carries little information on its own. This is a further
reason for treating the state-occupancy result with caution.

Occupancy counts bins wherever they fall in the mission. A source that
dips into the BL Lac-like band repeatedly but briefly and one that sits
there for a year return the same count, and only the second is a change
of state in any useful sense. We therefore also measure persistence.
\chg{For each source we record the longest run of consecutive bins that
lie $2\sigma$ below $\Gamma = 2.0$, using the bins with $TS \ge 25$ in
time order; with 30-day bins, a run of $n$ bins corresponds to at least
$30n$\,d.} We
then compare that run with the longest run obtained after randomly
shuffling the same source's bins \chg{200} times. Shuffling destroys
any ordering in time while leaving the distribution of $\Gamma$
untouched, so the comparison isolates temporal clustering from the mere
fact of spending time in the band.

\chg{We first compare the observed longest runs between the three
samples, before turning to the shuffling test.} The measurement requires
enough well-measured bins to define a run, and is made for 47 CLBs, 254
FSRQs and 144 BL Lacs. At 30-day cadence the median longest run is one
bin for CLBs, against zero for FSRQs and two for BL Lacs, and the
fraction of sources reaching three or more consecutive bins is 10.6,
3.5 and 30.6 per cent respectively. CLBs differ from both controls
($p = 1.1 \times 10^{-3}$ against FSRQs and $1.4 \times 10^{-4}$
against BL Lacs), and are again intermediate. They spend longer in the BL Lac-like band than FSRQs do, but nowhere near as long as BL Lacs. \chg{The longest such run in any CLB is six bins, so no CLB is
continuously BL Lac-like at this significance for more than about
half a year. Hard intervals of this length are expected from ordinary
jet variability; in OQ~334, for example, \citet{Ren2024b} identify
$\gamma$-ray epochs lasting from 4 to 767\,d.}

The shuffling test is the more informative of the two \chg{measures}. Only 8.5 per cent
of CLBs have a longest run that exceeds the shuffled expectation at the
5 per cent level, against 7.6 per cent of BL Lacs and 2.0 per cent of
FSRQs. \chg{The first two percentages, for CLBs and BL Lacs, are consistent with
the 5 per cent expected from chance alone, given the small numbers of
sources involved (4 of 47 CLBs and 11 of 144 BL Lacs).} In
other words, \chg{the bins in which a CLB is BL Lac-like are} not
clustered in time beyond what its own distribution of $\Gamma$ implies.
\chg{The fraction of bins in which a CLB is significantly BL Lac-like
($\Gamma$ more than $2\sigma$ below 2.0)} is real and larger than for
an FSRQ, but \chg{these bins are} scattered through the mission rather than gathered
into sustained intervals. This is the quantitative form of the
impression left by Fig.~\ref{fig:timeline}, and it is the measurement
that distinguishes a permanently intermediate source from one
alternating between two states.

\chg{A final check looks for secular evolution of the spectrum over the
mission.} Splitting each light curve
at the mission midpoint and comparing the mean photon index of the two
halves, the median absolute shift is 0.074 for CLBs, 0.067 for FSRQs and
0.052 for BL Lacs. CLBs are indistinguishable from FSRQs ($p = 0.61$)
and only marginally separated from BL Lacs ($p = 0.064$). Secular
evolution of the $\gamma$-ray spectrum over the mission is therefore no
more common in CLBs than in ordinary FSRQs.

\section{Central engine properties}
\label{sec:engine}

Every measurement presented so far -- the fractional variability of
Section~\ref{sec:fvar}, the spectral variability of
Section~\ref{sec:sigma}, the distribution shapes and power spectra of
Section~\ref{sec:dist}, the photon-index/flux relation of
Section~\ref{sec:indexflux} and the time-resolved behaviour of
Section~\ref{sec:time} -- describes the $\gamma$-ray emission, and
therefore the jet. It is useful to compare these with the properties of
the accretion flow, which can be estimated independently from optical
spectroscopy and so provides a check on the picture from a wholly
different observable.

\begin{figure}
\centering
\includegraphics[width=\columnwidth]{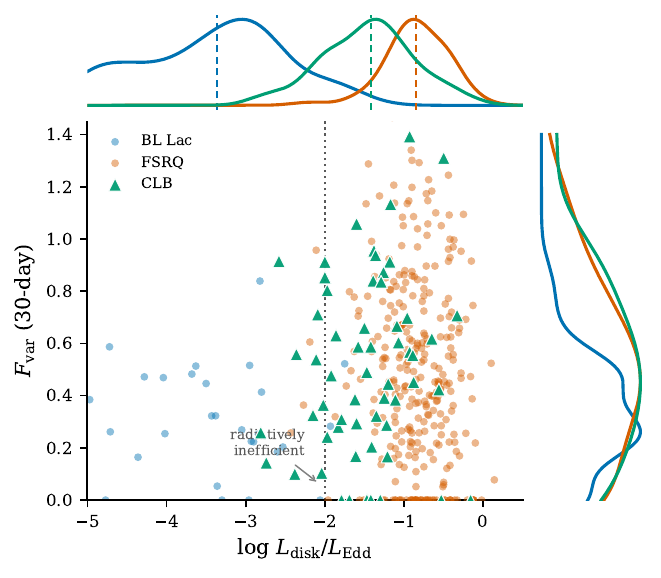}
\caption{Eddington-scaled disc luminosity against fractional variability
at 30-day cadence. Blue circles denote BL Lacs, orange circles FSRQs and
green triangles CLBs. The horizontal axis is derived from optical
spectroscopy \citep{Paliya2021} and the vertical axis from the
\textit{Fermi} light curves, so the two quantities are measured
independently of one another. Curves above and to the right are kernel
density estimates of the marginal distributions, with dashed vertical
lines marking the medians in $\log L_{\rm disk}/L_{\rm Edd}$. The dotted
vertical line at $-2$ marks the approximate boundary below which
accretion is expected to become radiatively inefficient. BL Lac values
are $3\sigma$ upper limits, since their broad lines are undetected, so
those points should be read as lying at or below the plotted position.}
\label{fig:accretion}
\end{figure}

Using black hole masses and disc luminosities from \citet{Paliya2021},
the median Eddington-scaled disc luminosity is
$\log L_{\rm disk}/L_{\rm Edd} = -1.42$ for the 61 CLBs with
measurements, $-0.85$ for 309 FSRQs and $-3.36$ for 29 BL Lacs. The
BL Lac values are upper limits, so the true separation is larger than
quoted. CLBs lie 0.57\,dex from the FSRQ median and 1.94\,dex from that
of the BL Lacs (Fig.~\ref{fig:accretion}). In other words, although CLBs
are formally intermediate, they sit close to one end of the interval
rather than in the middle of it.

Our CLB value is consistent with the $-1.59$ obtained by
\citet{Kang2024} from a partly overlapping sample, and with the
equivalent widths and broad-line luminosities reported by
\citet{Xiao2022}, from which most CLBs exceed the threshold associated
with efficient accretion. The accretion rates of CLBs are therefore
close to those of FSRQs, and the jet variability measured in the
preceding sections is consistent with that independent finding.

We also examined the ratio of the synchrotron peak luminosity to the
disc luminosity, as a measure of how strongly the jet dominates the
optical output. Taking the synchrotron peak fluxes of
\citet{Paliya2021} and the same cosmology as above, the median
$\log(L_{\rm syn,peak}/L_{\rm disk})$ is $+0.87$ for CLBs and $+0.17$
for FSRQs, a difference of $0.70$\,dex. This comparison should be
treated with caution, because CLBs are identified through the
equivalent width of their lines and a jet-dominated continuum reduces
that equivalent width, so the two quantities are not independent. The
difference is therefore at least partly a restatement of the selection,
and we report the measurement for completeness rather than as
independent evidence.

\section{Discussion}
\label{sec:discussion}

\subsection{Summary of the comparison}
\label{sec:summary}

\begin{figure}
\centering
\includegraphics[width=\columnwidth]{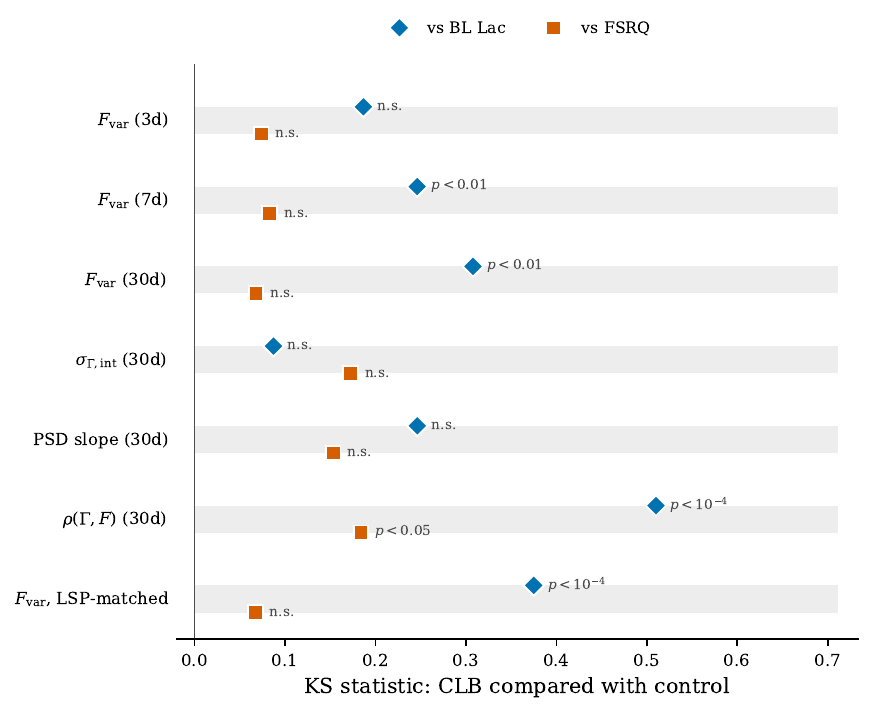}
\caption{Summary of the comparisons made in this paper. Each row is one
diagnostic. Blue diamonds give the KS statistic $D$ between the CLB
sample and the BL Lac control, and orange squares the same quantity
against the FSRQ control, so that distance from zero measures the size
of the difference rather than only its significance. The label beside
each point gives the corresponding probability; where the difference is
not significant (n.s.) at the 5 per cent level this is stated as such rather
than quoting a value, since the individual probabilities are given in
the relevant sections above. A row in
which the blue point lies well to the right while the orange point sits
near zero is one in which CLBs resemble FSRQs rather than BL Lacs. That
is the pattern in the fractional-variability rows. In the spectral
variability and power spectrum rows neither point is significant, since
those two diagnostics separate CLBs from neither control. The
photon-index/flux correlation is the one row in which both points are
displaced from zero.}
\label{fig:summary}
\end{figure}

Taken together, the diagnostics considered here show a strong overall
tendency for CLBs to resemble FSRQs rather than BL Lacs. The pattern is seen in the fractional variability, consistently across
the three cadences and after flux and sampling matching. It appears
again in the mean photon index, which places CLBs in the FSRQ band
rather than in the strip between the classes, and in the preferred
flux-distribution model, the centroid--width relation and the
Eddington-scaled accretion rate. The power spectral
density slope points the same way, the median for CLBs being 1.003
against 1.001 for FSRQs and 0.898 for BL Lacs, although the samples are
small and the separation is not significant. The noise-debiased spectral variability is uninformative rather than
supporting. It separates CLBs from neither control at 30-day cadence,
and its ordering of the three samples is not stable between cadences, so
it neither strengthens nor weakens the grouping. The per-source photon-index/flux correlation places CLBs between the two
canonical populations, as does the persistence of BL Lac-like spectral
states. Overall, these results favour an
FSRQ-like interpretation of the CLB population, while indicating that their photon-index/flux correlation, and the time they spend in the BL Lac-like band, retain an intermediate character.
Fig.~\ref{fig:summary} collects those comparisons that take the form of two-sample
tests on per-source quantities.

The way the results are distributed across the two comparisons is worth
drawing out. In Fig.~\ref{fig:summary} the orange points, which compare
CLBs with FSRQs, cluster near zero and are labelled as not significant
in almost every row. In this context a null result is not a weakness of
the measurement but the result itself. The comparisons against BL Lacs, made on the same data with the same
tests, do show significant differences in fractional variability, and
they survive both matching schemes. This confirms that the tests can
detect a difference of the relevant size when one is present. Where
neither comparison is significant, as for the spectral variability and
the power spectral density slope, the diagnostic simply does not
discriminate.

The absence of certain signatures is equally informative. Relative to
FSRQs, CLBs show no excess of bimodal flux distributions at any
cadence, which might be expected for sources alternating between two
emission states. Their excursions into the BL Lac-like spectral band are
no more clustered in time than a random reordering of their own bins
would produce. The direction of the optical
transition does not predict the sense of $\gamma$-ray spectral
evolution. And CLBs are no more spectrally variable than the control
samples.

Individual sources illustrate the same point. 4FGL~J2015.5$+$3710 is
recorded as an $F\!\rightarrow\!B$ transition in the optical, yet every
one of its usable $\gamma$-ray bins over 17.9\,yr lies in the FSRQ-like
band (Fig.~\ref{fig:lightcurves}). \chg{The spectrum on which its
BL Lac-like equivalent width is measured was published by \citet{Shaw2013},
within the period covered by the LAT, yet at no point in the mission does its
$\gamma$-ray index approach the BL Lac-like band.} Whatever changed in its optical
spectrum left no corresponding trace in its $\gamma$-ray spectrum.

\subsection{Interpretation}
\label{sec:interp}

The simplest reading of these results is that the changing-look
phenomenon leaves little imprint on the $\gamma$-ray jet. The jets of
CLBs vary like FSRQ jets, their spectra are distributed like FSRQ
spectra, and their central engines accrete at FSRQ-like rates.

This is what would be expected if the optical transitions arise from
variations in the jet continuum rather than from changes in the
accretion regime \citep{BlandfordRees1978, Ruan2014}. \chg{In this picture,}
the accretion flow and the jet remain in the same state throughout, and
the equivalent width of the broad lines changes because the continuum
against which they are measured changes. The classification then follows
the continuum, and no corresponding change need appear in the
$\gamma$-ray properties. \chg{The timescales agree: the BL Lac-like
intervals in CLB light curves last at most a few hundred days and are no
more persistent than chance allows (Section~\ref{sec:occupancy}), as
expected for a variable jet rather than for a change of accretion state
lasting years.} Two features of the sample fit naturally into
this picture. The first is the prevalence of low-synchrotron-peaked
sources among CLBs \citep{Xiao2022}, since these are the objects in
which the jet contributes most strongly at optical wavelengths. The
second is the position of CLBs in the accretion plane
(Fig.~\ref{fig:accretion}), close to the FSRQs and far above the regime
in which the broad-line region would be expected to disappear.

The time-averaged diagnostic that separates CLBs from FSRQs can be read
in the same framework. A source whose optical classification is set by its
continuum is one in which the jet contributes an unusually large
fraction of the optical output. If the jet component that dominates the
optical also contributes to the low-energy end of the $\gamma$-ray band,
a modest change in the way the spectrum responds to brightening would
follow. This interpretation should be treated as tentative. The effect
is small, the significance is modest, and with 64 sources the
measurement has limited precision. It is worth revisiting with a larger
sample.

An alternative reading of the same result is that CLBs are genuinely a
mixed population, containing both continuum-dominated FSRQs and a
smaller number of sources undergoing a real change of state. The
state-occupancy result of Section~\ref{sec:occupancy}, in which a
minority of CLBs occupy both spectral states, would be consistent with
this, as would the extended blue intervals visible for a few sources in
Fig.~\ref{fig:timeline}. Our data do not allow the two readings to be
separated, since both predict FSRQ-like population averages with a small
residual difference.

\subsection{Relation to previous work}
\label{sec:previous}

\citet{Kang2024} and \citet{Zhao2026} conclude that CLBs form a
transitional population, on the grounds that their time-averaged
properties are intermediate between those of the two classes. Our own
earlier analysis \citep{Malik2026} reached a compatible conclusion from
averaged $\gamma$-ray and X-ray parameters, finding CLBs between the two
classes but nearer the FSRQs. The present results refine that picture
rather than overturn it: when the same population is examined through
its light curves, the resemblance to FSRQs is closer than the averaged
parameters alone suggested. Our
measurement of the accretion rate agrees closely with theirs, and we
also find CLBs to be intermediate in the photon-index/flux relation. The
question is how the intermediate values are best read. A source that is
permanently intermediate and one that alternates between two states
produce the same average, so averaged quantities alone leave the choice open. The light curves add information that bears on
the choice, and they favour the first: CLBs resemble FSRQs throughout
the interval examined, rather than alternating between two states.

It is important to be clear about what this does and does not constrain.
Our measurements are sensitive to changes on timescales accessible
within a 17.9\,yr mission. They have no bearing on the much slower
evolution from FSRQ to BL Lac proposed on cosmological timescales
\citep{Bottcher2002, Cavaliere2002, Ajello2014}, which would be
unobservable here. The evolutionary interpretation advanced by
\citet{Kang2024} therefore lies outside the reach of these data. What our results
address is the narrower expectation that CLBs should presently display
intermediate or alternating behaviour in the jet, for which we find no
support.

Our results also sit comfortably alongside the detailed studies of
individual sources. Across the seven optically defined epochs of
OQ~334 between 2008 and 2019, \citet{Ren2024a} find a softer
$\gamma$-ray index in the FSRQ epochs than in the BL Lac epochs.
\citet{Tantry_2025} report index changes across flaring epochs of
S5~1803$+$784. Those analyses integrate over epochs chosen for each
source, which yields far greater precision per measurement than a
fixed-cadence light curve can provide. A source may therefore show a
clear spectral change across such epochs while contributing little to a
population-averaged comparison at fixed cadence. Our results should not
be read as arguing against those findings. For OQ~334 itself,
\citet{Mishra2021} found that the Mg\,\textsc{ii} line flux changed by
a factor of only 2--3 while the continuum varied far more.
\citet{2025MNRAS.541L..52C} found that its BL Lac-like $\gamma$-ray episodes
coincide with optical flares. Both favour the swamping interpretation
supported here.

\subsection{Limitations}
\label{sec:limits}

Several caveats apply to the results presented here.

\begin{enumerate}
\item The flux matching between CLBs and BL Lacs is imperfect ($p = 0.037$),
so a residual contribution from brightness to that comparison cannot be
excluded, although the corresponding comparison with FSRQs is well
matched and shows no difference.

\item At 3-day cadence the BL Lac control is reduced to 169 sources, with a
median of 56 usable bins per source, and the comparison is limited by sample size
rather than by any physical effect.

\item The LCR contains only sources already identified as variable, which
biases the BL Lac control towards its more variable members. This works
against the differences we report and so does not create them.

\item The distribution-model comparison is reported at all three cadences,
but the model choice is best determined at 30-day and progressively
noisier as the bins are shortened, since the mixture costs three extra
parameters while the number of bins per source falls. The shorter
cadences should therefore be read as supporting the 30-day result rather
than as independent of it. A hierarchical treatment, fitting the
population jointly so that sparsely sampled sources contribute to a
population-level mixture fraction without each having to be classified
on its own, would use the short-cadence data more efficiently than
per-source fitting does.

\item Tier~A transition directions are recorded without epochs, so the
directional test uses a trend over the full mission rather than a
comparison across the transition itself.

\item \chg{The CLB--FSRQ difference in the per-source photon-index/flux
correlation is significant at two of the three detection thresholds
tested (Section~\ref{sec:indexflux}) but not at the third, and should
be regarded as suggestive.}

\item Finally, the state-occupancy result of Section~\ref{sec:occupancy} is
not robust to the choice of thresholds and should be regarded as
suggestive only.
\end{enumerate}

\subsection{Future work}
\label{sec:future}

The most direct test of the continuum interpretation is a comparison
within individual sources rather than between populations. If the broad
lines are buried by the jet continuum, the optical continuum should be
brighter at the epoch when a source appears BL Lac-like than at the
epoch when its lines are visible. Such a test requires spectroscopic
epochs together with contemporaneous photometry. Cross-matching the
spectrum references of \citet{Paliya2021} against public optical surveys
suggests that a sample of order ten to twenty CLBs satisfies both
requirements, and we intend to pursue this separately.

Two further extensions follow from the limitations above. The first is a hierarchical treatment of the distribution models.
Fitting the population jointly, with the parameters of individual
sources drawn from a common distribution, would let sparsely sampled
sources contribute to a population-level mixture fraction without each
having to be classified on its own. That would use the 3- and 7-day
light curves more efficiently than the per-source fitting reported here,
on the timescales where \citet{Malik2025} find two-state structure most
visible.

The second is the photon-index/flux relation, the one diagnostic for
which CLBs differ from FSRQs. The limitation there is sample size rather
than method: with 64 CLBs the difference is measured at
$p = 0.042$, and the same measurement repeated on a few hundred sources
would establish or dismiss it. Progress therefore depends on the
confirmed CLB sample growing, which in turn depends on repeat
spectroscopy. Surveys such as LAMOST have already contributed
\citep{PenaHerazo2021}, and systematic searches continue
\citep{Kang2025}, so the sample can be expected to expand. Until it
does, this result should be regarded as a difference that is suggestive
but not firmly established.

\section{Conclusions}
\label{sec:conclusions}

We have characterised the $\gamma$-ray variability of changing-look
blazars using \textit{Fermi}-LAT light curves at three cadences, and
compared them with control samples of BL Lacs and FSRQs selected and
analysed in the same way. Our main conclusions are as follows.

\begin{enumerate}
\item CLBs and FSRQs have indistinguishable fractional variability, with
medians of 0.483 and 0.437 at 30-day cadence, while BL Lacs are
significantly less variable at 0.262. The result holds when the controls
are restricted to low-synchrotron-peaked sources and when they are
matched in flux and sampling.

\item The same grouping follows independently from the mean photon
index, the power spectral density slope, the preferred flux-distribution
model, the centroid--width relation and the Eddington-scaled accretion
rate, the last of which is derived from optical spectroscopy and so is
independent of the light curves. The noise-debiased spectral variability
separates CLBs from neither control and so does not discriminate between
the two pictures.

\item CLBs enter the BL Lac-like spectral band more often than FSRQs
do, but their runs of consecutive bins there are no longer than a random
reordering of their own light curves produces. The time they spend
looking like a BL Lac is scattered through the mission rather than
gathered into sustained intervals.

\item Relative to FSRQs, CLBs show no excess of bimodal flux
distributions at any of the three cadences. They show no excess spectral
variability against either control, and no correspondence between the
direction of the optical transition and the sense of $\gamma$-ray
spectral evolution.

\item The per-source photon-index/flux correlation places CLBs between
the two classes, with a median of $+0.149$ against $+0.077$ for FSRQs
and $+0.580$ for BL Lacs. This difference persists after flux matching, and is one of two
diagnostics, with the persistence of BL Lac-like states, on which CLBs
are displaced from the FSRQ sample.

\item Taken together, these results indicate that the changing-look
phenomenon leaves little imprint on the $\gamma$-ray jet. They favour an
interpretation in which the optical transitions arise from a variable
jet continuum rather than from a change of accretion regime, while
remaining compatible with a slow evolutionary transition on cosmological
timescales.
\end{enumerate}

\section*{Acknowledgements}

This work has made use of data and software provided by the
\textit{Fermi} Science Support Center, and in particular of light curves
from the \textit{Fermi}-LAT Light Curve Repository
\citep{Abdollahi2023}. We have also used the \textit{Fermi}-LAT Fourth
Source Catalog Data Release 4 \citep{Abdollahi2020, Ballet2023}, the
Fourth Catalog of Active Galactic Nuclei Data Release 3
\citep{Ajello2022}, the central-engine catalogue of \citet{Paliya2021},
and the changing-look blazar catalogue compiled by \citet{Kang2023}.
This research has made use of NASA's Astrophysics Data System and of the
NASA/IPAC Extragalactic Database. The analysis relied on
\textsc{numpy}, \textsc{scipy}, \textsc{pandas}, \textsc{astropy} and
\textsc{matplotlib}.

\section*{Data availability}

All data used in this work are publicly available. The $\gamma$-ray
light curves were obtained from the \textit{Fermi}-LAT Light Curve
Repository at
\url{https://fermi.gsfc.nasa.gov/ssc/data/access/lat/LightCurveRepository/}.
The 4FGL-DR4 and 4LAC-DR3 catalogues are distributed by the
\textit{Fermi} Science Support Center. The central-engine properties of
\citet{Paliya2021} are available at
\url{https://doi.org/10.5281/zenodo.4456365}, and the changing-look
blazar catalogue of \citet{Kang2023} at
\url{https://doi.org/10.5281/zenodo.8239456}. The derived quantities
underlying the figures and tables of this paper are available from the
corresponding author on reasonable request.



\bibliographystyle{mnras}
\bibliography{example} 





\bsp	
\label{lastpage}
\end{document}